\documentclass[fleqn,usenatbib]{mnras}

\usepackage{newtxtext,newtxmath}

\usepackage[T1]{fontenc}
\usepackage{ae,aecompl}

\usepackage{graphicx}	
\usepackage{amsmath}	

\usepackage{amssymb}	
\usepackage{threeparttable}
\usepackage{url}
\usepackage{color}

\newcommand{\SII}{[S~{\sc ii}]\ }
\newcommand{\OIII}{[O~{\sc iii}]\ }
\newcommand{\NII}{[N~{\sc ii}]\ }
\newcommand{\HII}{H~{\sc ii}\ }

\newcommand{\HI}{H~{\sc i}\ }
\newcommand{\Ha}{H$\alpha$\ }
\newcommand{\kms}{\,\mbox{km}\,\mbox{s}^{-1}}

\newcommand{\SIIHa}{[S~{\sc ii}]/H$\alpha$}
\newcommand{\NIIHa}{[N~{\sc ii}]/H$\alpha$}
\newcommand{\OIIIHb}{[O~{\sc iii}]/H$\beta$}

\newcommand{\pak}{PA_\mathrm{kin}}
\newcommand{\pap}{PA_\mathrm{phot}}
\newcounter{qub}

\newcommand{\revone}{}

\title[Ark~18 as result of dwarf-dwarf merger]{Search for gas accretion imprints in voids: II. The galaxy Ark~18 as a result of a dwarf-dwarf merger.}

\author[Egorova et al.]{
Evgeniya S. Egorova$^{1}$\thanks{E-mail: eshaldenkova@gmail.com},
Oleg V. Egorov$^{2,1}$\thanks{E-mail: oleg.egorov@uni-heidelberg.de},
Alexei V. Moiseev$^{3,1}$,
Anna S. Saburova$^{1,4}$,
\newauthor
Kirill A. Grishin$^{1,5}$,
Igor V. Chilingarian$^{6,1}$
\\
	$^{1}$ Lomonosov Moscow State University, Sternberg Astronomical Institute,
	Universitetsky pr. 13, Moscow 119234, Russia
	\\
	$^{2}$ Astronomisches Rechen-Institut, Zentrum f\"ur Astronomie der Universit\"at Heidelberg, M\"onchhofstr.\ 12--14, 69120 Heidelberg, Germany \\
	$^{3}$ Special Astrophysical Observatory, Russian Academy of Sciences, Nizhnii Arkhyz 369167, Russia\\
	$^{4}$ Institute of Astronomy, Russian Academy of Sciences, Pyatnitskaya st., 48, 119017 Moscow, Russia\\
	$^{5}$ Department of Physics, M.V. Lomonosov Moscow State University, 1 Vorobyovy Gory, Moscow, 119991, Russia\\
	$^{6}$ Center for Astrophysics -- Harvard and Smithsonian, 60 Garden Street MS09, Cambridge, MA 02138, USA
}

\date{Accepted XXX. Received YYY; in original form ZZZ}

\pubyear{2021}
\begin{document}
\label{firstpage}
\pagerange{\pageref{firstpage}--\pageref{lastpage}}
\maketitle

\begin{abstract}
The low-mass low-surface brightness (LSB) disc galaxy Arakelian~18 (Ark~18) resides in the Eridanus void and because of its isolation represents an ideal case to study the formation and evolution mechanisms of such a galaxy type. Its complex structure consists of an extended blue LSB disc and a bright central elliptically-shaped part hosting a massive off-centered star-forming clump. We present the in-depth study of Ark~18 based on observations with the SCORPIO-2 long-slit spectrograph and a scanning Fabry-Perot interferometer at the Russian 6-m telescope complemented by archival multi-wavelength images and SDSS spectra. Ark~18 appears to be a dark matter dominated gas-rich galaxy without a radial metallicity gradient. The observed velocity field of the ionised gas is well described by two circularly rotating components moderately inclined with respect to each other and a possible warp in the outer disc. We estimated the age of young stellar population in the galaxy centre to be $\sim$140~Myr, while the brightest star-forming clump appears to be much younger. We conclude that the LSB disc is likely the result of a dwarf--dwarf merger with a stellar mass ratio of the components \revone{at least} $\sim$5:1 that occurred earlier than 300~Myr ago. The brightest star forming clump was likely formed later by accretion of a gas cloud. 
\end{abstract}




\section{Introduction}

According to recent observational studies, the gas depletion time in present-day galaxies is 1--2~Gyrs \citep{Bigiel08,Bigiel2011,Leroy2008,Leroy2013}, so that the external gas accretion is needed to sustain star formation over longer periods of time \citep{Lilly2013, SA2014}. Several mechanisms of galaxy growth and gas replenishment are discussed, such as mergers \citep[e.g.,][]{LHuillier2012,DiMatteo2008}, galactic fountains \citep[e.g.,][]{Fraternali2006,Fraternali2008,Marinacci2010}, and cold gas accretion from filaments \citep{Semelin2005,Keres2005,Dekel2006,Dekel2009,Ceverino16}.

Voids are low-density regions in the large scale structure of the Universe. They are expected to have rich structure and include filaments and sheet-like substructures populated by low-mass dark haloes and galaxies \citep[e.g.,][]{vandeWeygaert2016,Cautun2014,VGS2013b}. Some observational evidence of filamentary alignments of galaxies in voids has also been recently revealed \citep[e.g.,][]{VGS2013,Chengalur17}. Moreover, numerical simulations suggest that cold accretion may still proceed along large-scale filaments onto low-mass haloes in voids \citep{Aragon13}.
\revone{The scenario of ongoing gas accretion was proposed in the course of Void Galaxy Survey \citep{VGS2012} for the galaxy VGS\_12 with \HI polar disk \citep{Stanonik2009} and the system of galaxies VGS\_31 \citep{VGS2013}. The inflow along the void filament was also discussed in \cite{Chengalur17} as a possible scenario for formation of the unusual extremely metal poor dwarf galaxy UGC3672A.
So, despite the low density environment, one can study interactions and merger events in voids, and also search for accretion of cold gas. Furthermore, the low rate of interactions makes it easier to distinguish among these processes.}

\revone{Recently} we started a project to study the processes of galaxy interactions, mergers and gas accretion in voids by compiling a sample of void galaxies of intermediate luminosity that reveal peculiar morphology or/and deviate from the reference ``metallicity--luminosity'' relation for Local Volume galaxies \citep{Berg12}. 
\revone{In our study we combine the data on ionised gas kinematics, optical and NIR morphology, and chemical abundances. The analysis of ionised gas kinematics allows us to reveal the regions with non-circular motions related to the stellar feedback, tidal disturbances, or accretion events. By comparing morphology and kinematics we can choose between different scenarios for the particular object (i.e., presence of a bar or multispin system, see Sec. \ref{sec:struct_and_kin} for details). Adding the data on metallicity distribution helps to identify the source of the disturbances, or about the source of external gas in the case of accretion \citep[e.g.,][]{SA2014,Lagos2018}.} 
More details on our sample and the results on the brightest galaxy in our sample, NGC~428 are presented in \cite{Egorova2019}.

In this paper we present detailed analysis for the low-mass galaxy Arakelian~18 \citep[Ark~18; ][]{Arakelian75}, \revone{while the study of other objects from our sample \citep{Egorova2019} will be presented in the forthcoming papers.}
Ark~18 is located in the Eridanus void \citep{Eridanus} at the distance $D\simeq 24.1$~Mpc\footnote{Calculated as $V_{LG}/73$, where $V_{LG}$ is the recession velocity in the Local Group coordinate system according to NED (\url{http://ned.ipac.caltech.edu/}). At this distance, the scale is 117~pc~arcsec$^{-1}$}.
\revone{It is considered isolated \citep{Karachentsev2011}, and has no known companions.} The deep optical images from SDSS Stripe~82 \citep{SDSSdr7,Stripe82,Stripe82_18} reveal blue low surface brightness flocculent spirals around a bright redder elliptically-shaped central part. 

Ark~18 is a low surface brightness (LSB) galaxy with a moderate mass and size of its disc. It is neither dwarf nor giant or ultra-diffuse galaxy (UDG), even though its LSB disc is only slightly more extended than the largest known UDGs. Similarly to giant low surface brightness galaxies (gLSBGs) it has prominent spiral structure but does not possess a very massive bulge. Another feature that makes Ark~18 similar to gLSBGs is that its  complex structure comprised of a high surface brightness galaxy embedded in an extended LSB disc \citep[see, e.g., ][]{ saburova2021,Saburova2019,lelli2010}. \revone{However, the LSB structure of Ark~18 has more moderate size in comparison to that of gLSBGs, its effective radius is  $\sim$5~kpc  ($>21$~kpc for gLSBGs). The effective radius of the inner component of Ark~18 is $\sim$1~kpc.}
 
\begin{figure}
    \centering
    \includegraphics[width=1.0\linewidth, trim={1.5cm 1.5cm 1.5cm 1cm}]{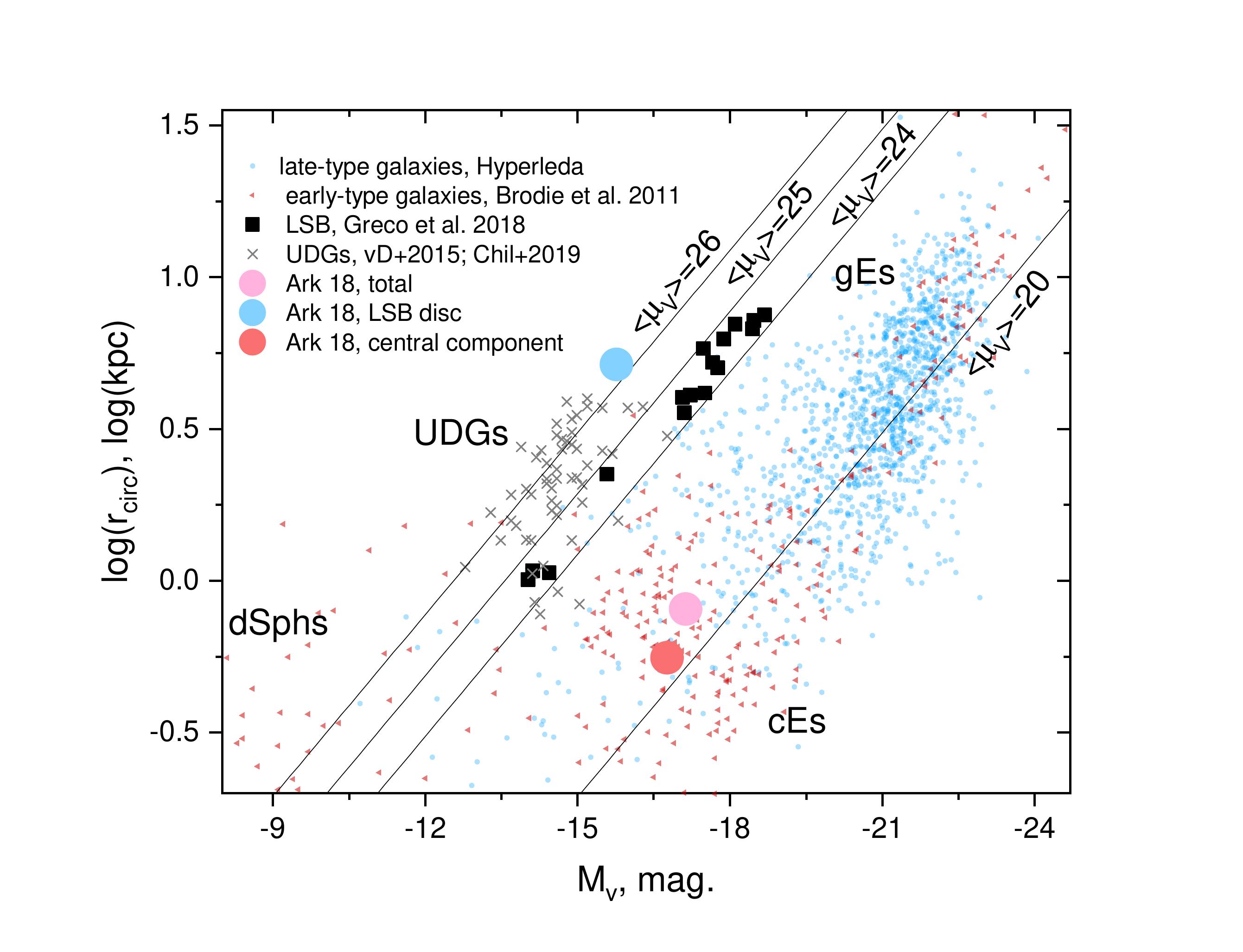}
    \caption{ The position of Ark~18 in the size versus luminosity relation (big circles). On the y-axis we plot the logarithm of circularized effective radius estimated following \citet{Greco2018}  $r_{circ}=(1-\epsilon)^{1/2}r_{eff}$, where $\epsilon$ is ellipticity and $r_{eff}$ is the radius containing the half of the luminosity. On the x-axis we give absolute V-band magnitude.    Red, pink and blue circles are related to the central component, the overall galaxy and LSB disc.  Black squares correspond to the position of LSB galaxies from \citet{Greco2018} for which the redshifts were available in Simbad database. Small \revone{red} triangles demonstrate the position of early-type galaxies from \citet{Brodie2011}. Small \revone{blue} circles correspond to the late-type galaxies (with $t$>2, \revone{where $t$ is the morphological type}) from the Hyperleda database. Grey crosses show the position of ultra diffuse galaxies from \citet{vanDokkum2015} and \citet{Chil2019}. The lines show the constant mean surface brightness.
    \label{fig:comparison}}
\end{figure}

We show the position of Ark~18 on the size--luminosity relation in Fig.~\ref{fig:comparison} in comparison to other early- and late- type galaxies, LSB galaxies, UDGs and compact ellipticals (cEs). \revone{As long as} the bright central component makes a bulk contribution to the luminosity of the system \revone{it defines the position of the whole galaxy on the diagram. Because we aim to understand not only the nature of the bright centre but also of the low-surface brightness structure of Ark~18}, we separately show the positions of the LSB disc, central component and the whole galaxy by circles of different colours. \revone{Here we do not consider that galaxies with different morphologies from the reference samples may also possess extended components with a low contribution to their total luminosities and compare the structural components of Ark~18 to their global properties.} We took the $V$-band magnitude for Ark 18 from Hyperleda\footnote{\url{http://leda.univ-lyon1.fr/}} \citep{hyperleda}, the bulge-to-disc luminosity ratio and effective radii were taken from our own analysis of its surface photometry. The inner component has the parameters which are close to those of dwarf early-type galaxies. At the same time, the LSB disc of Ark~18 lies close to the region of UDGs and typical LSB galaxies.
 
The blue colour of the spirals of Ark~18 indicates the presence of young stars which are formed from gas possibly accreted on-to it. Ark~18 is a gas-rich galaxy that is consistent with the external gas accretion scenario. 
Ark~18 was included in the HIPASS Bright Galaxy Catalog \citep{HIPASS_brightest} that contains 1000 H{\sc i} brightest galaxies compiled using HIPASS data from the southern sky. Its H{\sc i} mass, log$M_{HI}/M_{\odot}$ = 9.3 was calculated using the integrated H{\sc i} flux density from HIPASS \citep{HIPASS} and the distance adopted in this paper.
The sparse environment makes this galaxy ideal to study the formation of low surface brightness discs in low density environments.

The main properties for Ark~18 either taken from the literature or derived in this study are listed in  Table~\ref{tab:summary}. These include: the equatorial coordinates RA, Dec (J2000); the adopted distance $D$; the $B$-band apparent magnitude $m_B$ estimated from SDSS Stripe82 images (Sec.~\ref{sec:photometry}); the absolute magnitude $M_B$ corrected for the Galactic foreground extinction following \citet{Schlafly2011}; the star formation rate (SFR) (Sec.~\ref{sec:photometry}); the total mass of atomic hydrogen $\log(M_{HI})$; its ratio to the $B$-band luminosity $M_{HI}/L_B$; the systemic velocity $V_{sys}$ and both photometric and kinematic inclination $i$ and position angle $PA$ (Sec.~\ref{sec:struct_and_kin}); the oxygen abundance $12+\log\mathrm{(O/H)}$ (Sec.~\ref{sec:abundances}). 

The paper is organised as follows. In Section~\ref{sec:data} we describe our observations and data reduction. In Section~\ref{sec:results} we describe the performed data analysis and present its main results. In Section~\ref{sec:discuss} we discuss possible evolutionary scenarios of the galaxy in the light of our results and summarize our findings.

\begin{table}
	\centering
	\caption{Main properties and derived parameters of Ark~18. All values were obtained in this study unless otherwise noted}
	\label{tab:summary}
\begin{tabular}{l|c} \hline   \\ [-0.2cm]
Parameter & Value \\
\\[-0.2cm] \hline \\[-0.2cm]
RA	(J2000)$^a$	    & 00h51m59.62s   \\
Dec	(J2000)$^a$	    & -00d29m12.2s	\\
D$^a$, Mpc          & 24.1  \\
$m_B$           & 14.85  \\
$M_B$           & -17.2  \\
SFR, $M_\odot\ yr^{-1}$ & 0.1 \\
$\log M_{HI}^b$ 	& 9.3	\\
$M_{HI}/L_B$ 	& 2.3	\\
$i$ , deg 		    &  $67\pm1$ (inner$^c$) $58\pm3$ (outer$^d$)	\\
$\mathrm{PA}$, deg	    &  $358\pm3$ (inner$^c$) $344\pm5$ (outer$^d$) \\ 
$V_\mathrm{sys}$, $\kms$ &  1627$\pm9$	\\
$12+\log\mathrm{(O/H)}$ &  $8.20\pm0.04$\\
\\[-0.2cm] \hline \\[-0.2cm]
	\end{tabular}
\begin{tablenotes}
\item $^a$ from NASA/IPAC Extragalactic Database (NED)
\item $^b$ Derived using integrated \HI flux density from HIPASS \citep{HIPASS} and the distance adopted in this paper
\item $^c$ Derived from photometry, \revone{for $r<30''$}  
\item $^d$ Derived from kinematics, \revone{for $r>30''$}
\end{tablenotes}
\end{table}

\section{Observations and data reduction}
\label{sec:data}

\begin{table*}
\centering
\caption{Observation log}
\label{tab:Obs}
\begin{tabular}{llllllll} 
\hline
{Data set} & {Date of obs.} & {$T_{exp}$, s}& {$FOV$} & {\revone{Scale, arcsec~px$^{-1}$}} & {Seeing, $\arcsec$} & {\revone{Spectral range, \AA}} & {$\delta\lambda$, \AA} \\
{(1)} & {(2)} & {(3)} & {(4)} & {(5)} & {(6)} & {(7)} & {(8)}  \\  
\hline 
LS PA=145 & 2019 Nov 19 & 7200 & {$1\arcsec\times6.1\arcmin$} &  0.36 & 1.2 & 3650--7250 & 5.3  \\
LS PA=171 & 2019 Nov 20 & 6000 & {$1\arcsec\times6.1\arcmin$} & 0.36 & 2.5  & 3650--7250 & 5.3\\
LS PA=118 & 2019 Dec 19  & 10800 & {$1\arcsec\times6.1\arcmin$} & 0.36 & 2.5  & 3650--7250 & 5.3\\
FPI & 2016 Dec 23 & 11430 & {$6.1\arcmin\times6.1\arcmin$} & 0.71 & 2.4   & 8.7\AA\, around H$\alpha$ & 0.4 ($18 \kms$) \\ 
\hline
\end{tabular}
\end{table*}

Our observations were conducted in the prime focus of the 6-m telescope BTA of Special Astrophysical Observatory (SAO RAS) with the multi-mode focal reducer SCORPIO-2 \citep{SCORPIO2} in the modes of long-slit spectroscopy and 3D-spectroscopy with a scanning Fabry--Perot interferometer (FPI). The spectral data are complemented by multi-wavelength publicly available archival images, and also by SDSS spectra in the two regions of Ark~18. Below we describe each dataset in detail.

The log for BTA observations is presented in Table~\ref{tab:Obs}, which contains the information on (1) each spectral dataset including the position angle in case of long-slit observations; (2) the date of an observation; (3) the total exposure time $T_{exp}$; (4) the field of view (or width and length of the slit); (5) the pixel angular size of the obtained data; (6) the seeing quality that corresponds to the final angular resolution; (7) the available spectral range; and (8) the FWHM spectral resolution $\delta\lambda$. 

\subsection{Observations with Fabry--Perot interferometer}
\label{sec:fpi}

The observations in the \Ha emission line were performed with the high-resolution scanning FPI providing the spectral resolving power of about $R\sim16000$ in a free spectral range between the neighbouring interference orders of $8.8$\,\AA. The operating spectral range around the \Ha emission line was cut by a narrow bandpass filter eliminating the contamination from the \NII lines around H$\alpha$. The galaxy was exposed at two different position angles in order to remove the \revone{ghost images}, as described in \citet{Moiseev2008}. During the observations we have consecutively obtained 40 interferograms for each field orientation at different gaps between the FPI plates. The exposure for each individual channel was 150~s, however 19 channels for second field orientation were exposed for only 120~s each.

The two datasets were reduced separately using the software package running in the \textsc{idl} environment \citep{Moiseev2002,Moiseev2015}. After primary reduction, air-glow lines subtraction, photometric and seeing corrections using reference stars, wavelength calibration, the individual wavelength channels were combined into data cubes, where each pixel in the field of view contains a 40 channel-long spectrum around the red-shifted \Ha emission line. Then both cubes were co-added with removing the ghosts and artifacts. During the observations the atmospheric seeing quality was 1.1--1.8~arcsec. The final spatial resolution after all smoothing during the data reduction process is about 2.4~arcsec.

H$\alpha$ line profiles were analysed by fitting a single-component Voigt profile, which is a good representation of the emission line convolved with the FPI instrumental profile \citep{Moiseev2008}. The fitting procedure yields the flux, line-of-sight velocity, and velocity dispersion (free of the instrumental broadening) for each spatial element of data cube. In the final maps we have masked the regions with the low signal-to-noise ratio ($S/N<3$). \revone{The typical error of the velocity estimation for this $S/N$ corresponds to $\sim9\kms$ and drops to $\sim2\kms$ at $S/N=10$ \citep{Moiseev2015}.}  As a result, we obtained a deep image in the \Ha line revealing a large number of star-forming clumps in the LSB disc of Ark~18 (see Fig.~\ref{fig:slitpos}). The resulting maps for other parameters are analysed in Section~\ref{sec:struct_and_kin}. The obtained \Ha image originally was not in energy units. To perform the calibration, we used the \Ha map obtained by \cite{GildePaz2003} at the Palomar telescope as a reference. We found a linear regression between the flux values for the pixels bright in \Ha in our and their maps and extrapolated it to the whole map. \revone{The uncertainty of the obtained conversion factor is about 15 per cent that is suggested further as our relative error for the derived \Ha fluxes and SFR (this does not include possible uncertainties of the calibration of the reference \Ha map used).}

\subsection{Long-slit spectroscopic observations}

\begin{figure}
    \centering
    \includegraphics[width=\linewidth]{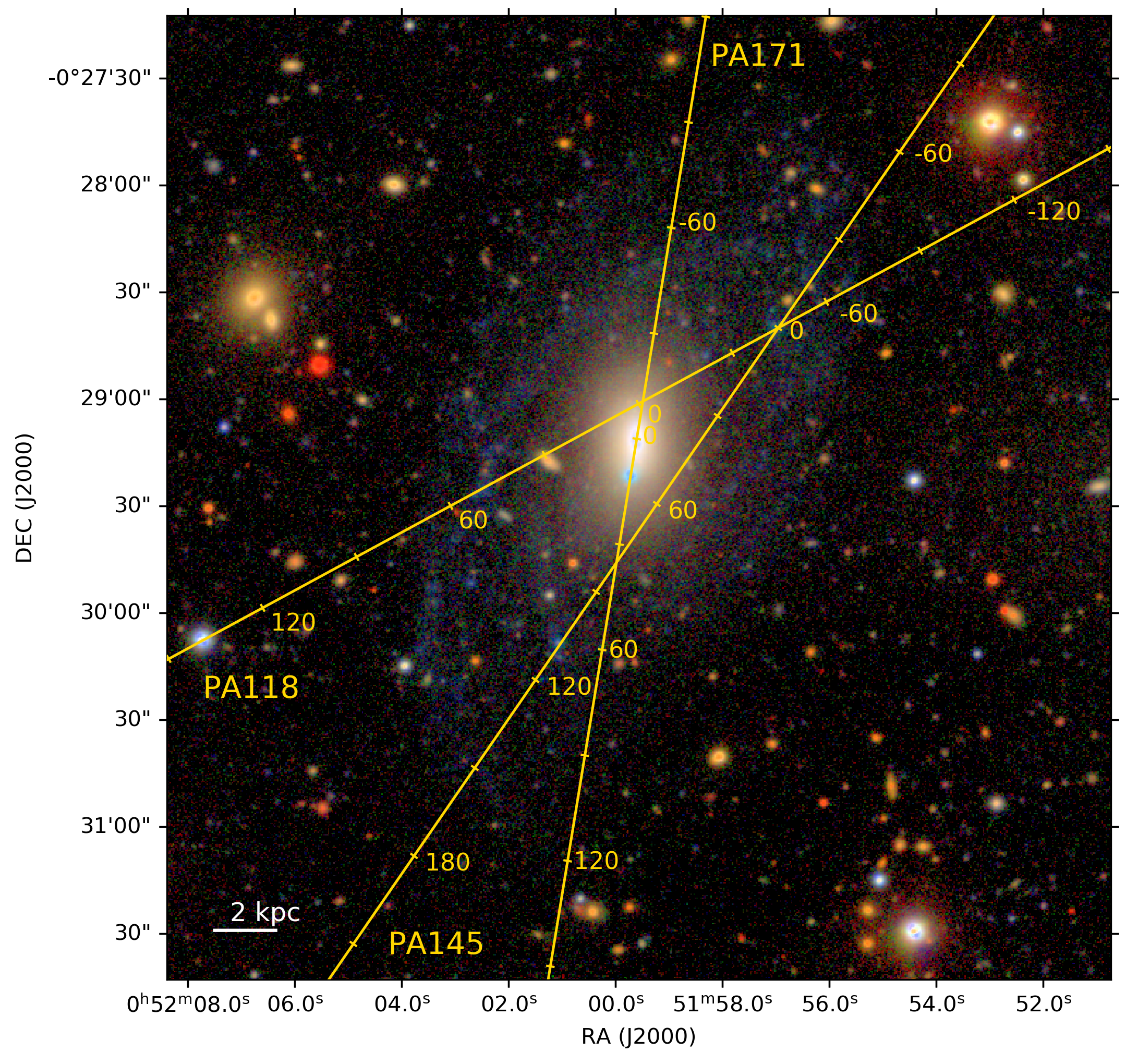}
    
    \includegraphics[width=\linewidth]{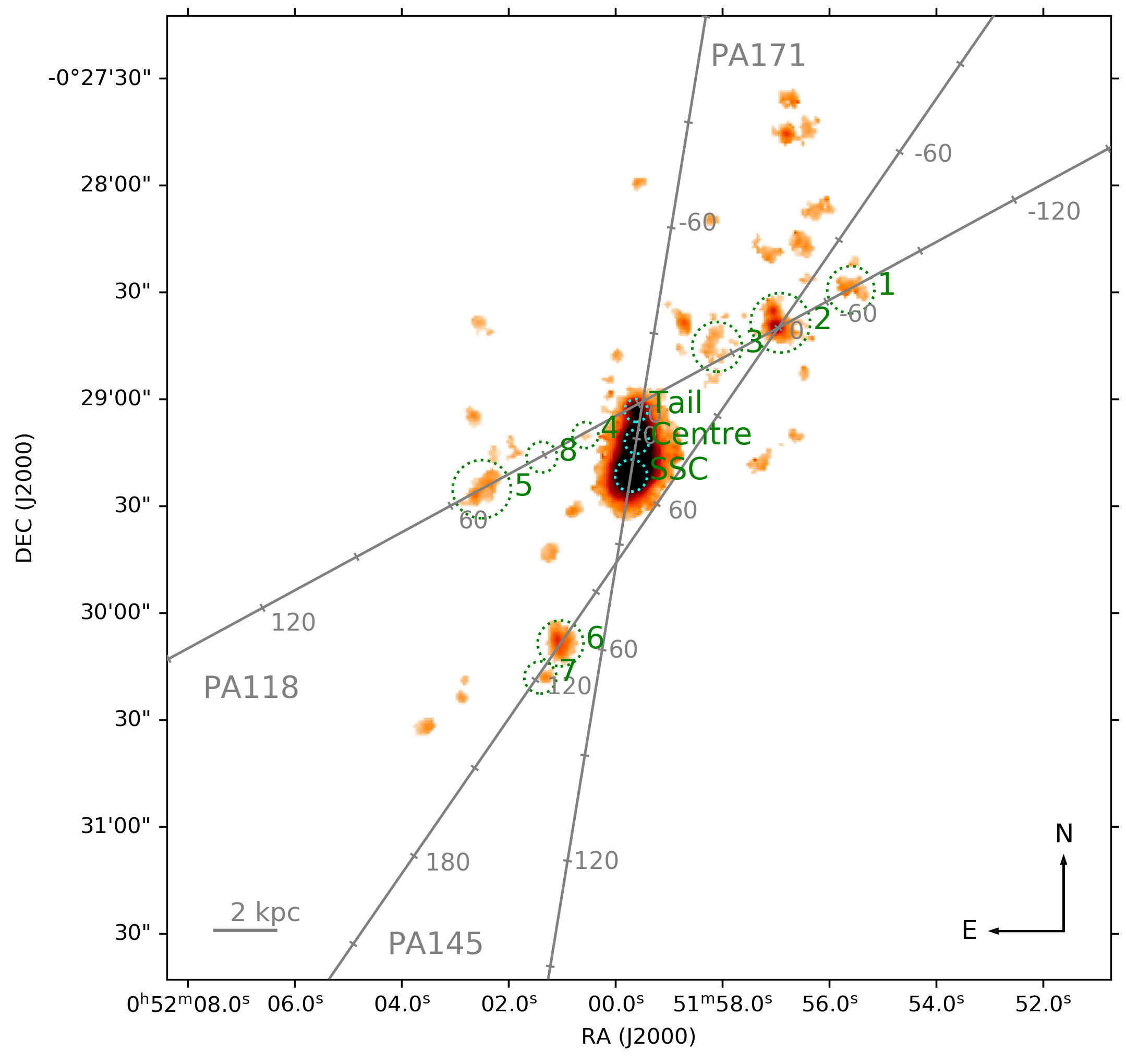}
    \caption{Slit positions overlaid on the SDSS Stripe 82 $g$ (blue), $r$ (green), $i$ (red) false-colour image (\textbf{top panel}) and on the masked H$\alpha$ map obtained from our FPI observations (\textbf{bottom panel}) shown in the same spatial scale. The \HII regions discussed in the text are highlighted by green circles. \revone{The regions denoted by digits are located in the LSB disc, while the ``Tail'', ``Centre'' and ``SSC'' regions are from the central component of the galaxy.}}
    \label{fig:slitpos}
\end{figure}

We obtained three spectra at different slit positions. \revone{Slit positions overlaid on an SDSS Stripe~82 color image and on the H$\alpha$ map are shown in Fig.~\ref{fig:slitpos}.} One of them with $PA=171^\circ$ crosses the galaxy centre and the brightest off-centre star-forming clump \revone{(hereafter denoted as the super star cluster or the SSC)}. The two other slit positions ($PA=118^\circ$ and $PA=145^\circ$) cross some of the brightest star-forming clumps in the LSB disc which we detected in our FPI observations. The slit $PA=118^\circ$ also passes through a relatively bright \Ha `tail' near the central part of the galaxy having the minimal measured gas velocity dispersion \revone{(see Fig.~\ref{fig:FPI})}. \revone{These data are used for analysis of gas excitation and chemical abundances (see Section~\ref{sec:emission}).}

For our observations we used the volume phase holographic grism VPHG1200@540, which covers the wavelength range 3650-7250~\AA\, with a typical spectral resolution of 5.3~\AA\ (as estimated from the \textit{FWHM} of air-glow emission lines). The slit width was 1~arcsec.

The data were reduced in a standard way using a pipeline written in {\sc python} for SCORPIO-2 long-slit data. The main steps of the data reduction process include bias subtraction, line curvature and flat-field corrections, wavelength calibration, and air-glow line subtraction. Each individual exposure (5--9 for each PA) were reduced separately and then combined with cosmic-ray rejection. The spectra were calibrated to the wavelength scale using the He-Ne-Ar lamp reference spectrum obtained during observations. A set of LED lamps inside the integration sphere was used as a flat field source (see the detailed description of the used calibration system in \citealt{Afanasiev2017}). One of the spectrophotometric standards (either BD+25d4655 or BD+28d4211) was observed at a close airmass immediately before or after the science target, and was used for the absolute flux calibration. After the initial data reduction, we fitted the processed spectra taking into account the parameters of the line-spread-function (LSF) of the spectrograph resulted from the fitting of the twilight sky spectrum observed during the same observing runs.

We convolved high-resolution PEGASE.HR~\citep{LeBorgneetal2004} simple stellar population models (SSP) with the instrumental LSF and fitted them against fully reduced spectra of the galaxy. For this purpose we used the full spectral fitting technique \citep{Chilingarian2007a, Chilingarian2007b}, which allows us to fit a spectrum in a pixel space and returns parameters of stellar kinematics and stellar population simultaneously in the same non-linear $\chi^2$ minimization loop.

\begin{figure}
    \centering
    \includegraphics[width=\linewidth]{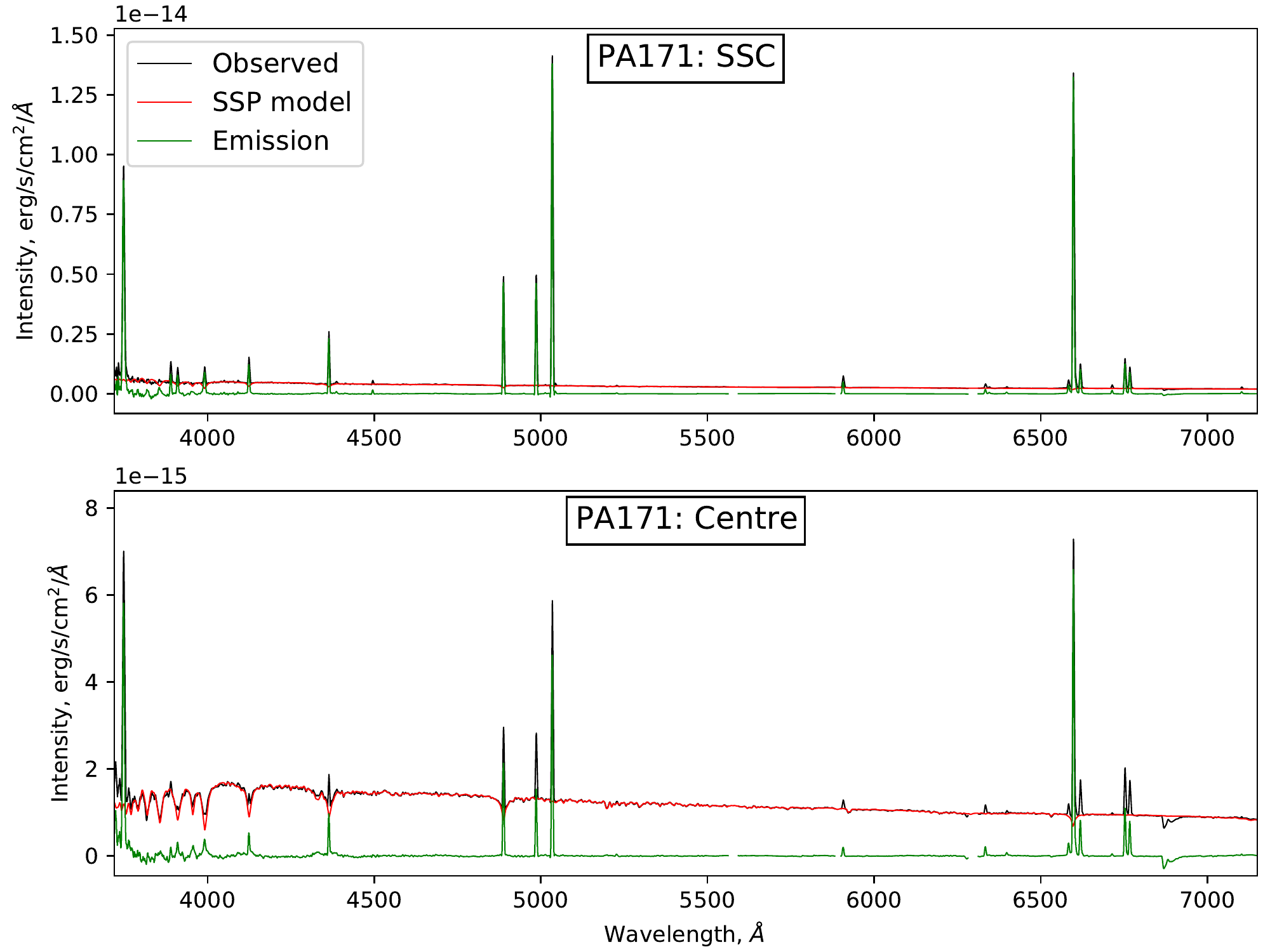}
    \caption{\revone{Integrated spectra of the regions ``SSC'' (top panel) and ``centre'' (bottom panel) extracted within the bounds given in Table~\ref{tab:spectral_par_bulge} demonstrating the quality of the recovery of emission-line spectra (green) after subtraction of the best-fitting stellar population model (red) from the observed spectra (black). }}
    \label{fig:spec_example}
\end{figure}

The emission-line spectra were obtained by subtracting the best-fitting stellar population models from the observed spectra \revone{(an example for two bright central regions are shown in Fig.~\ref{fig:spec_example})}. After that we fitted Gaussian profiles convolved with the instrumental LSF into emission lines to estimate their fluxes and line widths. We used our software package in the \textsc{idl} environment and based on the \textsc{mpfit} \citep{mpfit} non-linear minimization routine. All measured flux ratios were corrected for reddening based on the derived Balmer decrement using the reddening curve from \cite{Cardelli1989} parametrized by \cite{Fitzpatrick1999}. When the observed ratio of H$\alpha$/H$\beta$ was below the theoretical value 2.86, we did not apply any reddening correction (thus assuming $E(B-V)=0$~mag). To estimate the final uncertainties of the measured line fluxes, we quadratically added the errors propagated through all data-reduction steps to the uncertainties returned by \textsc{mpfit}.

\subsection{Archival datasets}

To perform surface and aperture photometry we used Sloan Digital Sky Survey \citep{SDSSdr7} 
Stripe~82 \textit{u, g, r, i}-band images. Stripe~82 is a $2.5^\circ$--wide region along celestial equator, that has been repeatedly scanned 70--90 times in \textit{u, g, r, i, z} filters as a part of the SDSS. IAC Stripe~82 Legacy Project \citep{Stripe82,Stripe82_18} provides deep co-added images that are 1.7--2.0 mag deeper than single-epoch SDSS images. The reduction and non-aggressive sky subtraction were aimed to preserve information about low surface brightness features.

The two available SDSS spectra (Plate=394, MJD=51913, FiberID=182 and Plate=692, MJD=52201, FiberID=240) were used to analyse stellar population of the galactic centre and of the brightest star-forming clump. Their analysis is described in Section~\ref{sec:sed}.

For spectral energy distribution (SED) fitting we also used available ultraviolet data in the $FUV$ and $NUV$ bands from Galaxy Evolution Explorer satellite \citep[GALEX; ][]{GALEX} and infrared IRAC 3.6~$\mu$m band from the Spitzer Space Telescope \citep{Spitzer_Spies}. For the high-surface brightness parts of Ark~18 we also used data from the VISTA Hemisphere Survey \citep{2013Msngr.154...35M} in the $J$ and $K_s$ bands.

\section{Results}\label{sec:results}

\subsection{Morphology, star formation rate and structural properties}
\label{sec:photometry}

\begin{figure*}
    \centering
    \includegraphics[width=0.9\linewidth]{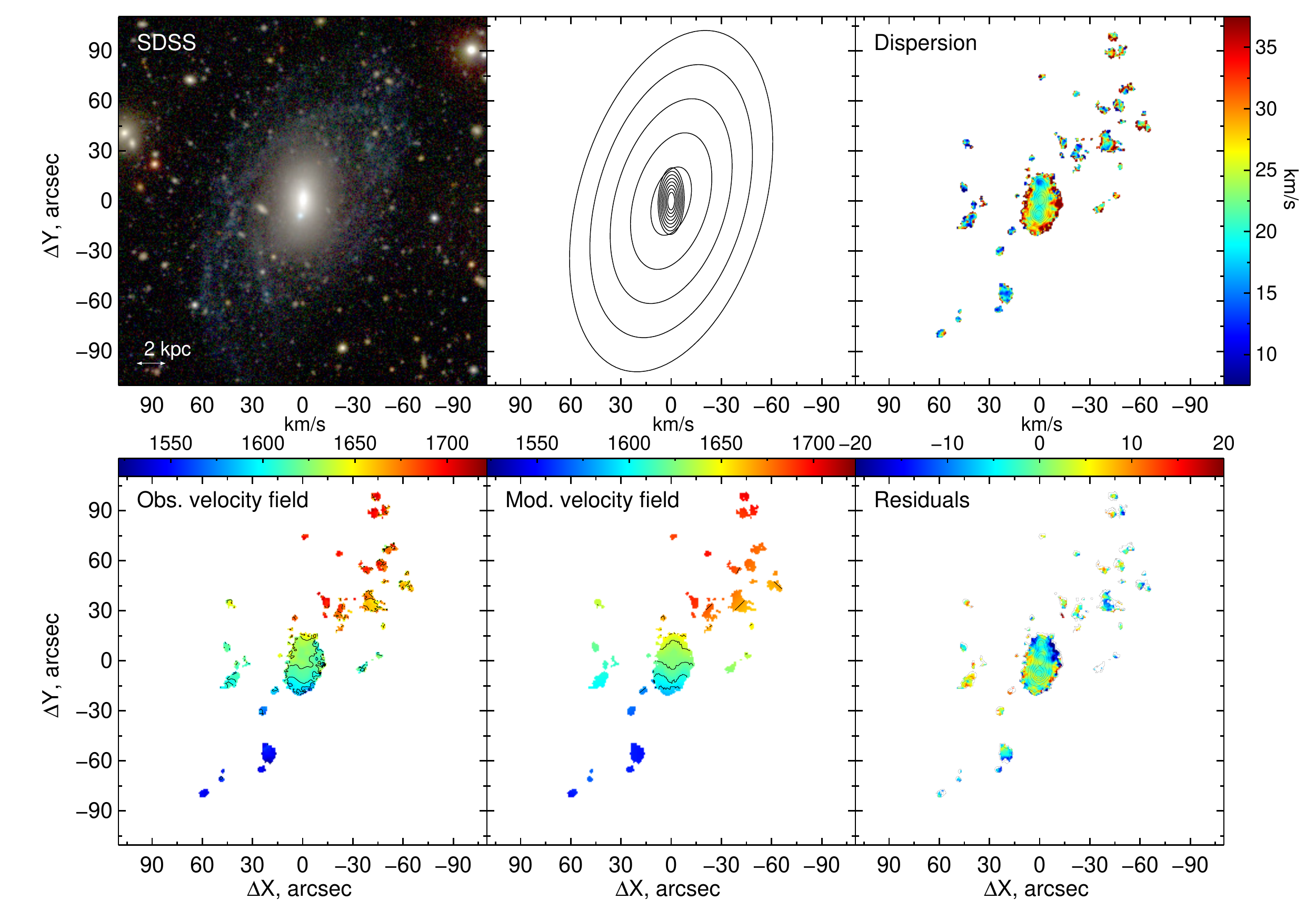}
        \caption{\textbf{Top row:} an SDSS Stripe82 $gri$-image (left-hand panel); \revone{the accepted orientations of the inner and outer discs} (middle panel); the velocity dispersion of ionised gas, \revone{with the H$\alpha$ brightness levels overlaid} (right-hand panel). \textbf{Bottom row:} the observed velocity field of ionised gas (left-hand panel); \revone{modelled velocity field} (middle panel); the velocity residuals after subtracting a model velocity field from the data, \revone{with the H$\alpha$ brightness levels overlaid}  (right-hand panel). \revone{The areas with S/N<3 were masked on four latter panels.}}
    \label{fig:FPI}
\end{figure*}

A false-colour $gri$-image constructed from SDSS Stripe 82 data (Fig.~\ref{fig:slitpos}, top) reveals an LSB disc with a prominent blue flocculent spiral structure surrounding a redder dense central bulge-like component. A bright blue clump of star formation, ``the SSC'' is clearly visible on the southern edge of the dense central \revone{component}. This SSC is the brightest region of star formation in the galaxy that contributes about 55 per cent of the total \Ha emission from the central dense part. The second brightest \HII region with a contribution of about 26 per cent is located in the photometric centre of the galaxy (hereafter referred to as ``the centre''). 

Our FPI map in the \Ha line (Fig.~\ref{fig:slitpos}, bottom) shows that many low-luminosity \HII regions are sitting in the LSB disc at different galactocentric radii. They are located within the blue spiral arms.  We estimate the \Ha flux from LSB disc to be about $1.7\times10^{-14}$~erg~s$^{-1}$~cm$^{-2}$, while the total measured \Ha flux from the galaxy is $1.9\times10^{-13}$~erg~s$^{-1}$~cm$^{-2}$. Hence, the LSB disc contributes at about 9 per cent to the total \Ha luminosity of Ark~18 (not counting possible regions overlapping with the \revone{central component}). The observed \Ha fluxes correspond to the total star formation rate $SFR\simeq0.1\ M_\odot\ yr^{-1}$ adopting the conversion from \citet{Kennicutt1998}. The corresponding SFRs for the SSC and the centre are $\sim0.05$ and $\sim0.023\ M_\odot\ yr^{-1}$, respectively. 

To the North of the centre of Ark~18 there is an extended structure of ionised gas with an \HII region at its end which we hereafter refer to as ``the tail'' (its structure is better visible in Figs.~\ref{fig:FPI} and \ref{fig:lsresults_pa171}). This tail is clearly bent with the distance from the galaxy centre. It is almost indistinguishable in SDSS images, however well noticeable in \Ha maps despite its rather low brightness (it contributes to about 3 per cent of the total \Ha flux). Towards the tail we observe the minimal value of the \Ha velocity dispersion $\sigma_{\mathrm{gas}}$ in the central part of Ark~18 and no any peculiarities in velocity field (see Fig.~\ref{fig:FPI}). We discuss the origin of the $\sigma_{\mathrm{gas}}$ increase in the central \HII regions outside the tail further in Section~\ref{sec:broadcomp}

We estimated the apparent magnitude in $B$ band provided in Table~\ref{tab:summary} from the aperture photometry of SDSS Stripe 82 $g$- and $r$-band images, using the conversion into Johnson bands by \cite{Lupton05}. Apparent background and foreground objects were masked when we performed the aperture photometry.

To investigate how the photometric inclination and position angle of Ark~18 change with the radius we performed the isophotal analysis of \revone{SDSS Stripe 82} $g$, $r$, and $i$-band images using the \textsc{iraf}\footnote{\textsc{iraf}: the Image Reduction and Analysis Facility is distributed by the National Optical Astronomy Observatory, which is operated by the Association of Universities for Research in Astronomy, Inc. (AURA) under cooperative agreement with the National Science Foundation (NSF).} \textsc{ellipse} task. Background and foreground objects along with bright SF regions of Ark~18 were masked out. We fitted WISE \citep{WISE} 3.4$\mu$m and Spitzer 3.6$\mu$m data in a similar fashion. The derived profiles of inclination and position angle are consistent with each other, except for the very central part where WISE and Spitzer images are affected by poorer resolution. For the subsequent analysis we use the values obtained from the SDSS Stripe~82 data because they are significantly deeper. In Section~\ref{sec:struct_and_kin} we compare the radial variations of position angle derived from the isophotal analysis with that from internal kinematics. 

We performed a two-dimensional image decomposition of the \revone{ less affected by dust and bright star-forming clumps SDSS Stripe 82} {\it i-}band image of Ark~18 with {\sc galfit} \citep{galfit} using a two-component model that included an inner Sersic component and an outer exponential disc. We obtained the following parameters of the components: \revone{effective radius $r_e=6.98 $~arcsec, effective surface brightness  $\mu_e=20.76$~mag arcsec$^{-2}$, Sersic index $n=1.69$ for the central component; effective radius $r_e= 39.25$~arcsec, effective surface brightness  $\mu_e= 25.59$~mag arcsec$^{-2}$, Sersic index $n= 1$ for the disc (it was fixed during the decomposition).}  \revone{According to the criteria defined by e.g. \citet{Kormendy2011}, the inner component is a pseudo-bulge because of its low Sersic index\footnote{We obtained even lower values of $n\sim 1.4$ from the decomposition of the surface brightness profiles obtained from the isophotal analysis} $n<2$, it has active star formation and does not look rounder than the disc. At the same time, high bulge-to-total luminosity ratio observed in Ark~18 is not typical for pseudo-bulges  \citep{Kormendy2011}.} Since the formal statistical uncertainties returned by {\sc galfit} are known to be significantly underestimated \citep[see, e.g.][]{Zhao2015} we do not provide them here. To verify the reliability of our image decomposition we performed a one-dimensional fitting of azimuthally averaged surface brightness profile, \revone{obtained with \textsc{ellipse} task for SDSS Stripe 82 {\it i}-band image}. The results turned to be consistent within 10 per cent for the central component and 23 per cent for the LSB disc. These values hence can be used as uncertainties of the best-fitting parameters.

\subsection{Global kinematics of the ionised gas}
\label{sec:struct_and_kin}

\begin{figure}
    \centering
    \includegraphics[width=0.95\linewidth]{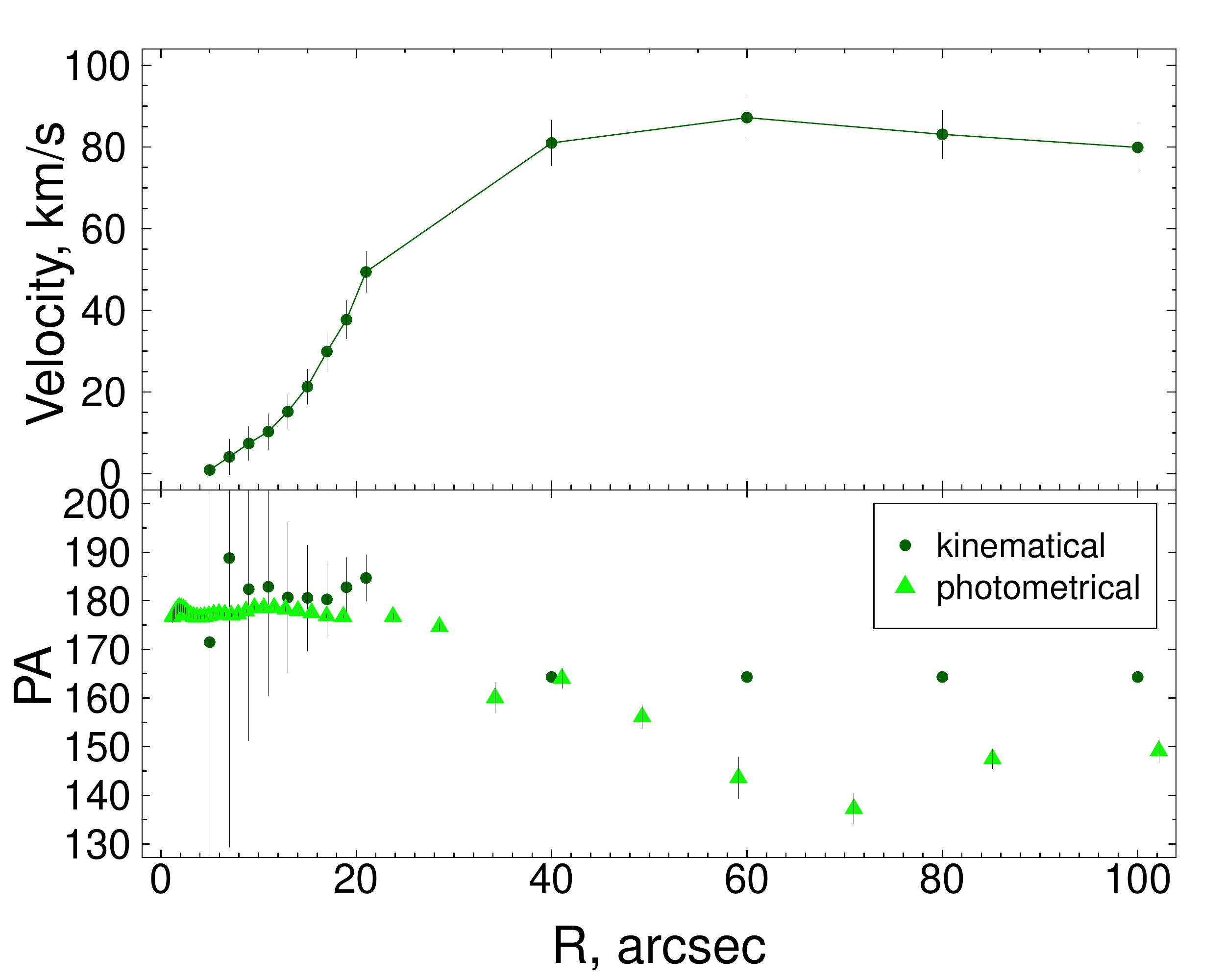}
    \caption{\textbf{Top:} The deprojected rotation curve for Ark~18 obtained from the observed ionised gas velocity field using tilted-ring analysis. \textbf{Bottom:} The distribution of position angles along the major axis. The values obtained from the analysis of the \Ha velocity field are marked by dark green circles (tilted-ring results for $r<30''$ and a flat disc model for the outer region). The values derived from isophotal analysis of Stripe 82 $r$-band image are marked by light green triangles.}
    \label{fig:PA}
\end{figure}

In Fig.~\ref{fig:FPI} we show the maps of the line-of-sight (LOS) velocity and velocity dispersion ($\sigma_{\mathrm{gas}}$) distributions derived from the \Ha FPI data cube. 
The rotation of the LSB disc is obvious from the velocity field, and it clearly rotates in the same sense as the central bright part of the galaxy, however the amplitude of the LOS velocity is not as large (about $20-30 \kms$) in the centre. The \Ha velocity dispersion in the galaxy is typical for that observed in other dwarf galaxies \citep{MoisKlypin2015}.

According to our isophotal analysis, the central region of Ark~18 has a position angle of the major axis $PA\approx-2\degr$ (at $r<30''$) while the outer disc has $PA=150-160\degr$ (i.e. $330-340\degr$) and also a lower ellipticity compared to the inner component. This difference will suggest the following two possible interpretations of the observed structure of Ark~18:

\begin{enumerate}
\item A low surface brightness disc with a central stellar bar. In such a case, we should see a turn of inner isovelocity lines due to non-circular (radial) motions reflecting gas streaming \citep[e.g.][]{Lindblad1996,Lin2013}. 
\item A multispin system with \revone{matter} rotating in the two different planes for the inner and outer regions like in polar rings or warped galaxies \citep[e.g.][]{Sparke2009}.
\end{enumerate}

The analysis of a two-dimensional \revone{ionised gas} velocity field and subsequent comparison of the radial behaviour of kinematic ($\pak$) and photometric ($\pap$) position angles of the major axis allows us to distinguish between these two scenarios. 
\revone{In the case of a disc with stellar bar we expect the turn of $\pak$ in the direction opposite to the major axis of the inner isophotes $\pap$, whereas for multispin system they should be in good agreement with each other. Indeed, in the former case we are dealing with non-circular motions under action of asymmetrical gravitational potential, while the latter scenario implies circular motions in the different planes  \citep[see ][and references therein]{Moiseev2000,Moiseev2004}}.
For example, such analysis revealed non-circular motions related to a bar in several other galaxies from our sample (\citealt{Egorova2019}, Egorova et al. in preparation).

We analysed the H$\alpha$ velocity field with the ``tilted-ring'' technique  \citep{Begeman1989A} modified for the studies of ionised-gas velocity fields of dwarf galaxies as described in \citet{Moiseev2014}. First, we have fixed the coordinates of the centre for Ark~18. The velocity field was then sliced into narrow elliptical rings in agreement with the mean values of inclination $i_0$ and position angle $PA_0$ estimated from the parameters of the optical isophotes (Sec.~\ref{sec:photometry}). 
\revone{As an initial approximation of $i$ and $PA_{kin}$ in the ``multispin'' approach we used two sets of $i_0$ and $PA_0$ (separately for the inner and outer disc), whereas only the outer disc orientation was used in the ``bar'' approach.}
In each ring, the following parameters of a quasi-circular rotation were determined: $\pak$, rotation velocity $V_{rot}$, and systemic velocity $V_{sys}$. On the next iteration, $V_{sys}$ was fixed at the mean value along the radius. The inclination of circular orbits was also kept fixed at $i_0$, because the non-circular gas motions in the inner part and a poor emission-line filling of the outer disc prevented the direct estimation of the $i$ from the gas kinematics.

The both assumptions for a galaxy structure gave the same conclusion: the value of $\pak$ in the inner region ($r<30''$) are in a good agreement with $\pap$, i.e. it is significantly (on 15--25 deg) twisted from the outer disc $PA_0$. Such radial behaviour of $\pak$ is impossible for the gas motions in the presence of a gravitational potential of a bar, where $\pak$ should twist in the direction opposite to the major axis of the inner isophotes $\pap$ \citep{Moiseev2000,Moiseev2004}.

\begin{figure*}
    \centering
\includegraphics[angle=0,width=0.45\linewidth]{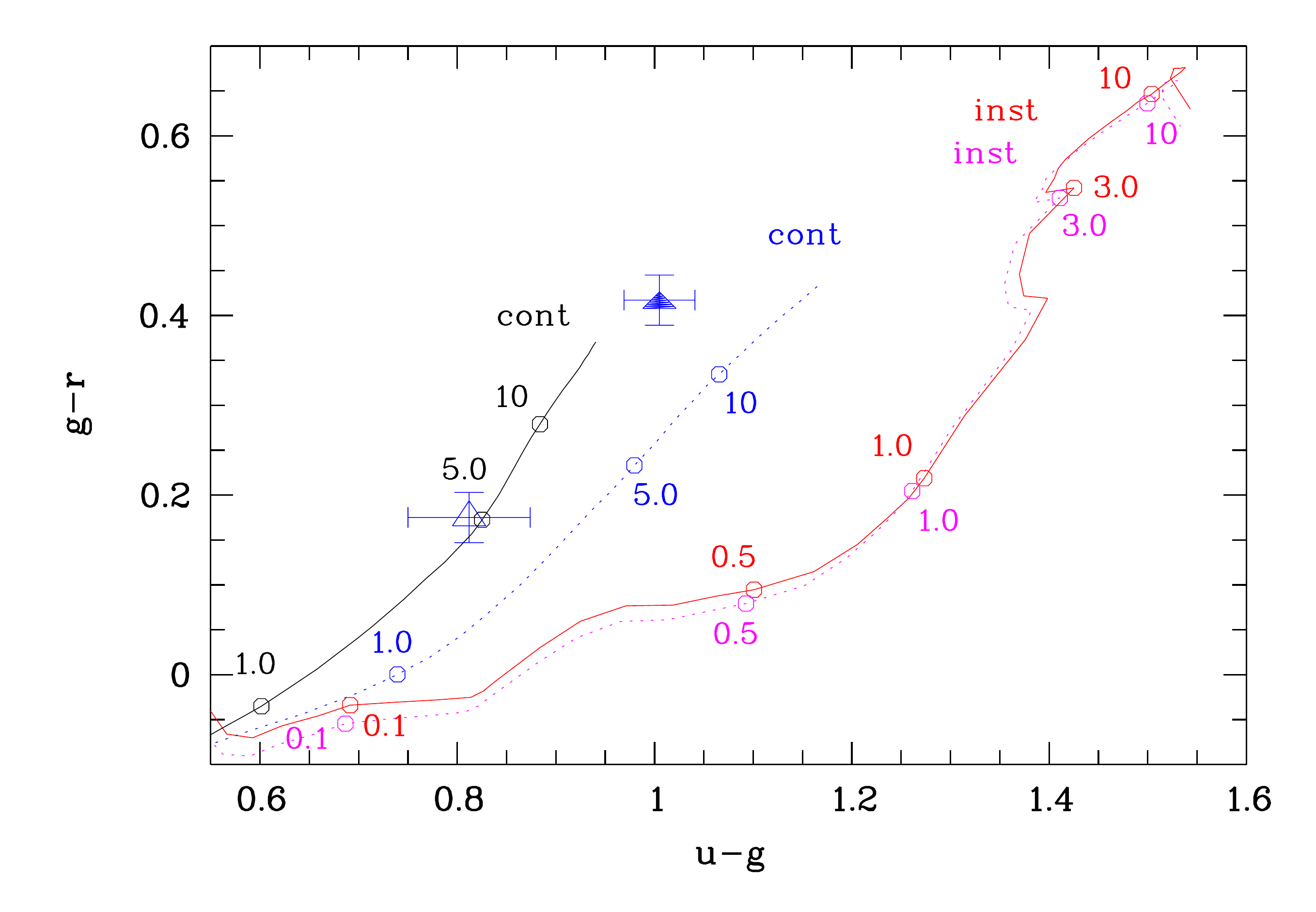}
\includegraphics[angle=0,width=0.45\linewidth]{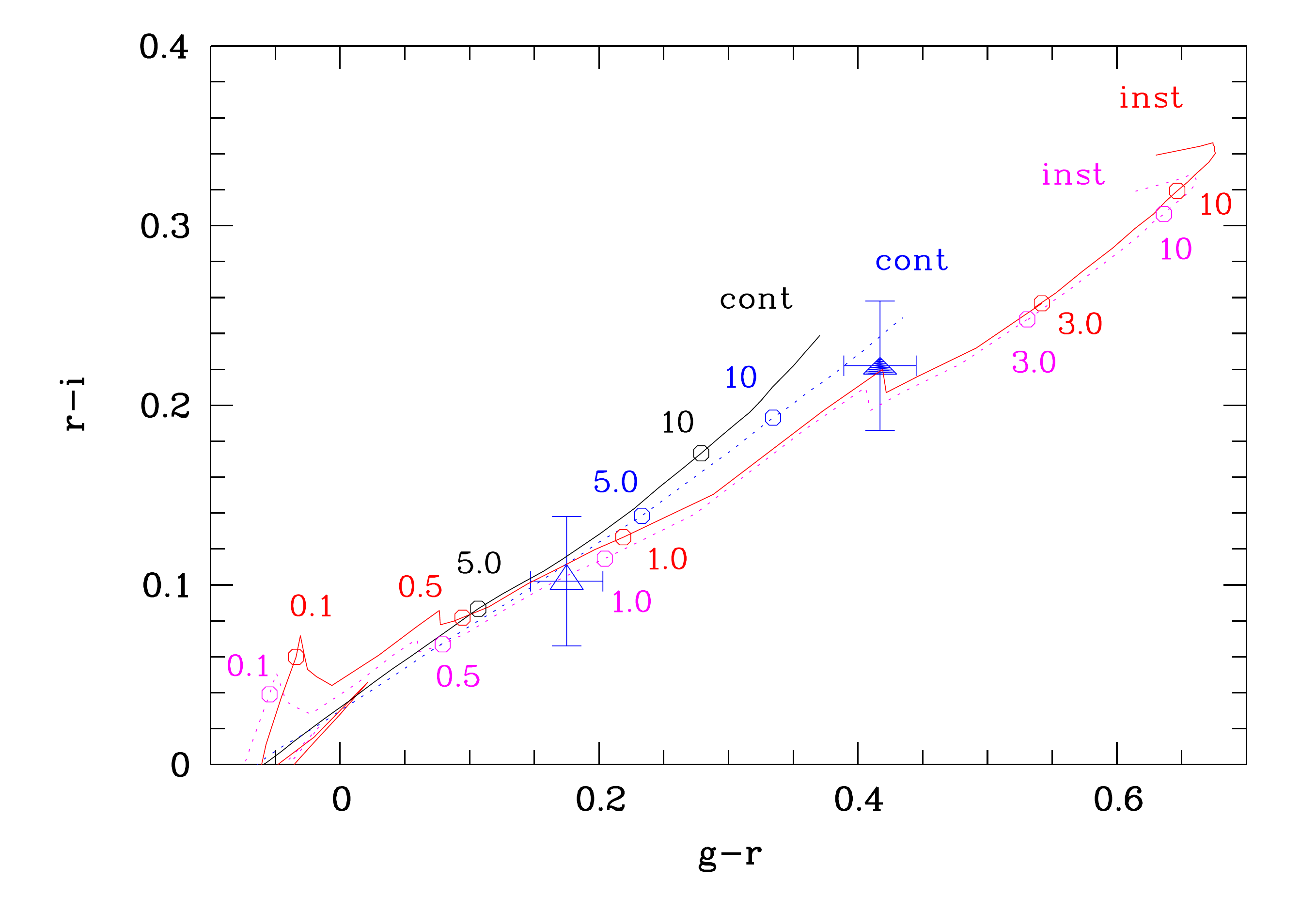}
    \caption{Colour--colour diagrams. \textbf{Left panel}: $(g-r)$ vs $(u-g)$; \textbf{Right panel:} $(r-i)$ vs $(g-r)$. Extinction-corrected colours of the central \revone{component} (blue filled triangles) and the low surface brightness periphery (blue empty triangles) of Ark~18 are overlaid on evolutionary tracks computed with {\sc pegase.2} \citep{PEGASE2} for metallicity Z = 0.004 (instantaneous SF for two initial mass functions (IMF): Salpeter \citep[][the red solid line]{Salpeter1955} and Kroupa \citep[][the red dashed line]{Kroupa02}; continuous star formation for the same two IMFs: Salpeter (the black solid line) and Kroupa (the black dashed line)). The values along the tracks are in Gyr.}
    \label{fig:colors}
\end{figure*}

Therefore, the ``multispin'' assumption is much more preferable than a high-contrast compact bar nested in the LSB disc. In order to better estimate the kinematic parameters of the outer disc we fitted the observed velocity field at $r=\revone{30}-100$~arcsec with a ``flat disc'' model having the constant value of $\pak$ and $i$ and several points  with different $V_{rot}$ along the radius \citep{Finkelman2011}. The resulting rotation curve and the radial behaviour of $PA_\mathrm{kin}$ together with the isophotal $\pap$ are shown in Fig.~\ref{fig:PA}. \revone{The model velocity field (tilted-ring for the inner disc and a flat model for the outer one) and} residual velocities after the model subtraction are presented in Fig.~\ref{fig:FPI}. Also, this figure shows the accepted orientation of the gas circular orbits in the inner and outer disc based on the orientation parameters presented in Fig.~\ref{fig:PA} and Table~\ref{tab:summary}.

As a rotation centre we accepted the ``centre'' \HII region that coincides with the photometric centre of the galaxy in the SDSS $i$ band. It also coincides with the centre of symmetry evaluated from the velocity field of the inner disc. 
We also considered the SSC star forming clump as dynamical centre for either inner or outer disc. The later assumption gives us similar conclusions about gas kinematics, however with larger spread of $\pak$ in the central region and the difference in $V_{sys}$ values between inner and outer discs on about $15\kms$. Hence, we conclude that the SSC is offset from the dynamical centre of the galaxy.

The $PA$ and $i$ values presented in Table~\ref{tab:summary} give the value of the angle between the inner and outer discs in the range $\Delta i=$20--30 degs, depends on the mutual spins orientation \citep[see equation (1) in][]{Moiseev2008arp}. Therefore in Ark~18 we observe an outer warped rather than a polar disc. Unfortunately, the \Ha emission filling factor of the outer disc is too poor to study a smooth radial  change of the disc's inclination and position angle. To accurately describe the disc structure in this galaxy, a deeper high-resolution \HI mapping is required. 

\subsection{Stellar populations of Ark~18}
\subsubsection{Stellar age from broad-band colours}
\label{sec:colors}

We analysed the colours of Ark~18 galaxy using {\sc pegase.2} package evolutionary tracks \citep{PEGASE2}. To estimate the colours we performed the aperture photometry for central bulge-like region and outer low surface brightness disc in SDSS Stripe 82 in $u, g, r, i$-band images outside bright SF regions. After that we compared the colours $u-g$, $g-r$, $r-i$, corrected for the Galactic extinction, with synthetic evolutionary tracks for the metallicity Z=0.004, close to the estimates obtained for Ark~18 (see Sec.~\ref{sec:abundances}). The tracks for a continuous SF law with a constant SFR and the instantaneous starburst were used as the two \revone{extreme variants of SF scenarios}, both with the two initial mass functions, Salpeter \citep{Salpeter1955} and Kroupa \citep{Kroupa02}. The corresponding plots are shown in Fig.~\ref{fig:colors}.

From Fig.~\ref{fig:colors} it is clear that the colours of the inner \revone{component} (excluding the very central part and the SSC) are rather red and correspond to a model for continuous SF with time elapsed from the onset of star-formation $\sim$13-14~Gyr (i.e. old stellar population), or more likely to the mixture of old and young stellar populations. At the same time, the \revone{central component} lies in the ``blue cloud'' occupied by the galaxies with normal star formation history (SFH) in 3D parameter space $g-r$ vs $NUV-r$ vs $M_r$ \citep{Chilingarian2012}.
The colours of outer low surface brightness disc (excluding relatively bright SF regions) correspond to significantly younger ages of stellar population, $\sim$5~Gyr for continuous SF. Because of the mixture of stellar population of different ages is very probable in our case, a full spectral energy distribution (SED) fitting is desirable, which we present in the next subsection of the paper.

\subsubsection{SED fitting and stellar mass}\label{sec:sed}

\begin{figure*}
    \centering
    \includegraphics[width=0.86\linewidth]{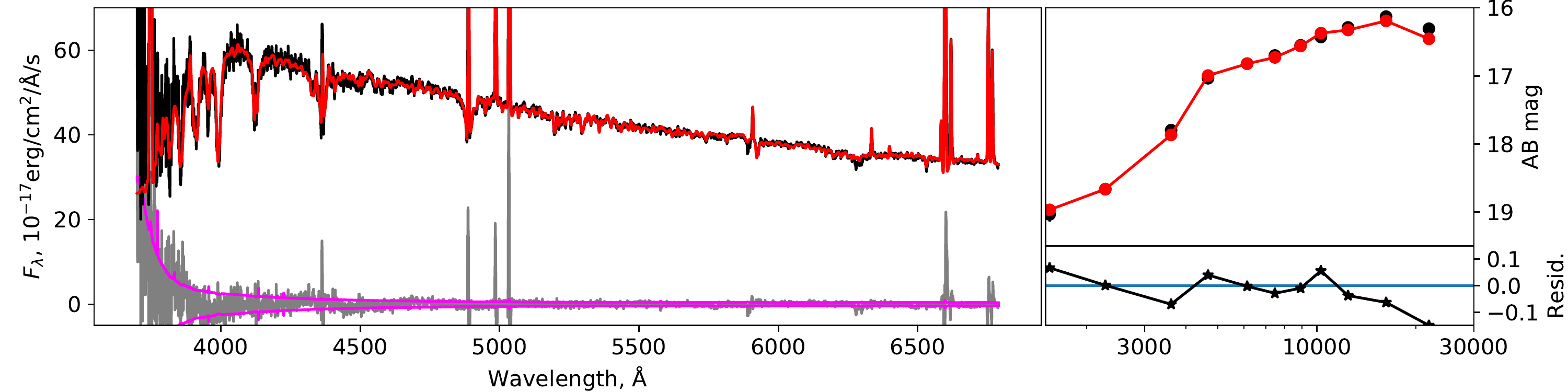}
    \caption{The SDSS spectrum for the ``centre'' region of Ark18 in $F_{\lambda}$ units \revone{(left panel, black)} and its SED in AB magnitudes (\revone{right panel}, black), the best-fitting model spectra and SED (red), flux uncertainties (purple) and residuals (grey line and black asterisks). The SED fitting residuals are shown in the bottom-right panel.}
    \label{fig:nbursts_central_spec}
\end{figure*}

\begin{figure*}
    \centering
    \includegraphics[width=0.9\linewidth]{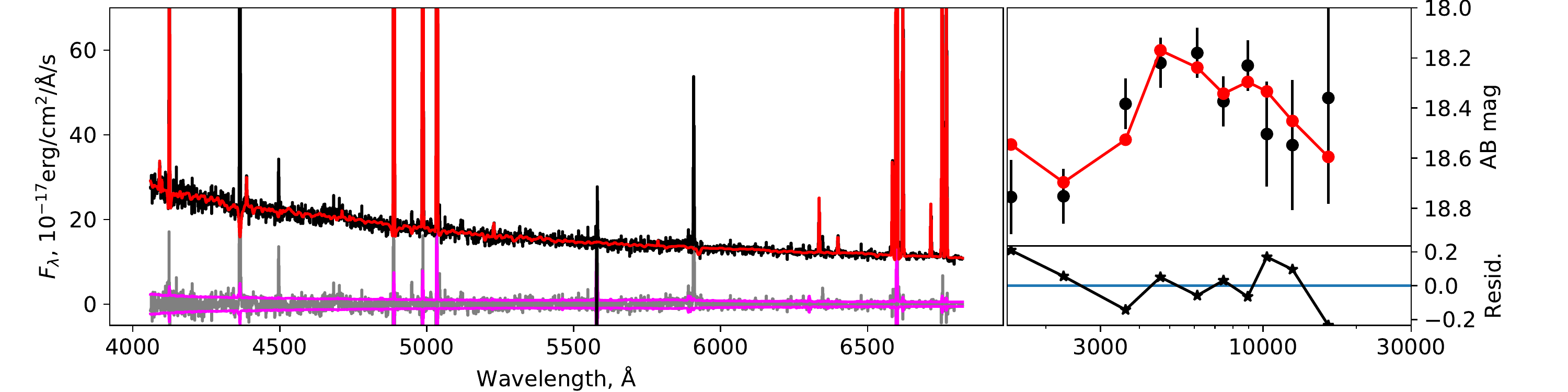}
    \caption{The SDSS spectrum for the ``SSC'' region of Ark18 in $F_{\lambda}$ units and its best-fitting models. The symbols are the same as in Fig.~\ref{fig:nbursts_central_spec}}
    \label{fig:nbursts_clump_spec}
\end{figure*}

\begin{figure}
    \centering
    \includegraphics[width=0.9\linewidth]{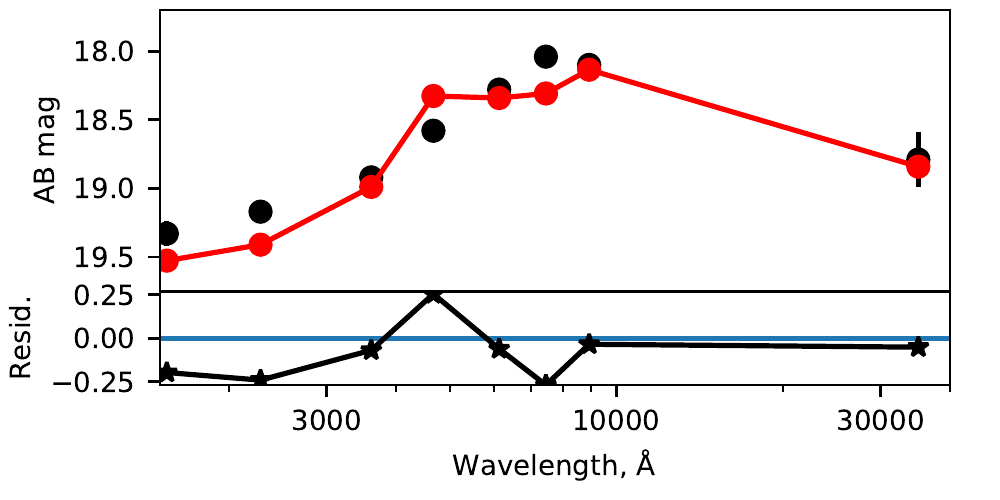}
    \caption{A broad-band SED (black circles) for the LSB disc of Ark~18 composed of broadband fluxes from far-UV (GALEX) to near-IR (Spitzer). The best-fitting model (red) is a SSP with the age of 127~Myr and metallicity fixed to $-0.5$~dex. The residuals are shown by black asterisks in the bottom panel. }
    \label{fig:sed_one_ssp}
\end{figure}

To analyse stellar populations in the ``centre'' region we simultaneously fit with the {\sc Nbursts+phot} \citep{Nburstsphot} code an SDSS optical spectrum and an SED composed of broadband fluxes from far-UV to mid-IR extracted from the same aperture. We apply aperture corrections described in \citet{2017ApJS..228...14C} to GALEX and Spitzer data having relatively poor spatial resolution. The observed SED cannot be adequately explained by a single burst of star formation and even by two starbursts because the UV part significantly deviates from the model. To improve the agreement, in our analysis we use the model by \cite{2019arXiv190913460G} with a truncated constant star formation from the age of 13~Gyr till some moment, where the starburst happens followed by the SFR cut-off to zero. The model include self-enrichment in the framework of a ``leaky box'' model with the delayed iron enrichment analytically formulated in \citet{2018ApJ...858...63C}. As an input grid of models we use MILES-based model spectra \citep{2015MNRAS.449.1177V} and broadband fluxes generated with the {\sc pegase.2} code. Because of a relatively low spectral resolution of SDSS data and young stellar populations with relatively featureless spectra, we can only reliably estimate stellar radial velocity but not the velocity dispersion \citep{2020PASP..132f4503C}.

During the fitting procedure we use a model with the following components fitted simultaneously as a linear combination: (i) a starlight model convolved with a Gaussian LSF and a Gaussian LOSVD and (ii) a model of emission lines with profiles described by another Gaussian LOSVD. The linear combination is then multiplied by a 9-th degree polynomial continuum to account for imperfections of flux calibration in both models and an observed spectrum.

We obtained the following parameters for the ``centre'' region: the age of truncation $t_{\mathrm{trunc}}=137.8\pm1.8$~Myr, the fraction of consumed gas 0.50$\pm$0.02, the mass fraction of stars born in the final starburst 20~per~cent, a linear coefficient of galactic winds $\lambda=1.5$. A slight disagreement in the Spitzer IRAC1 band might be the result of intense PAH lines not included in our pure stellar population model. 
The best-fitting model overlaid on the observed SDSS spectrum of Ark~18 and its SED is shown in Fig.~\ref{fig:nbursts_central_spec}. The model is consistent with the results obtained in previous Section~\ref{sec:colors}, a mixture of old and young stellar population.

For a spectrum and SED modelling of the SSC region we followed a similar procedure as for ``centre'' region, but we utilized the extension of {\sc  MILES} SSP models for ages younger than 63 Myr. Emission lines in SSC region are so intense that they contribute up to 30 per cent to the total fluxes in the $g$ and $r$ bands. We corrected broad band fluxes in $g$ and $r$ for emission lines contribution based on their flux estimated from the SDSS spectrum (i.e. running the fitting procedure twice). The modelling yielded a mean stellar age of $t_{\emph{SSP}} = 7.76 \pm 0.25$~Myr, and the stellar metallicity $[Z/H] = 0.01 \pm 0.08$ dex. For the model shown in Fig.~\ref{fig:nbursts_clump_spec} we assumed the value of $E(B-V) = 0.07$~mag as derived from the Balmer decrement (see Table~\ref{tab:spectral_par_bulge}), but in-detail analysis of $\chi^2$ behaviour for different $E(B-V)$ values showed that $\chi^2$ reaches its minimum for $E(B-V)=0.12$~mag. Bright emission lines in the loci of the prominent absorption lines minimize a statistical contribution of the spectrum to the estimates of stellar population properties making an SED the main source of the age and metallicity information. 

The very low surface brightness of the LSB disc did not allow us to perform the stellar population analysis based on full spectrum fitting. However, we were able to model the broad-band SED of the disc (see Fig.~\ref{fig:sed_one_ssp}). The SED was composed of broadband magnitudes extracted from the elliptical aperture at the LSB disc region with foreground and background sources masked. We modelled the extracted SED with one SSP with the metallicity fixed to $-0.5$~dex corresponding to the average value for the disc gas phase metallicity obtained from the emission line analysis (see below). From the best fitting results we estimate the age of the stellar population in the LSB disc to be between 60 and 500~Myr with the best-fitting value of about 130~Myr. This estimate however should be treated with caution and rather as an indication of significantly younger age of stellar population of LSB disc because of the low signal subject to systematics related to sky subtraction and, consequently, the poor quality of the modelling (Fig.~\ref{fig:sed_one_ssp}). Also it is worth noticing that from this fit we cannot draw any conclusion on the presence of possible underlying old stellar population.

We estimated stellar masses of both components of Ark~18 (central part and LSB disc) using magnitudes obtained from the {\sc galfit} decomposition. For the central component we used its total magnitude in the $i$ band and $M/L_{*,i} = 0.39 \revone{\pm 0.05}\ (M/L)_{\odot}$ obtained from the SED fitting results. \revone{As we mentioned above, for LSB-disc the quality of our broad-band SED fitting was poor. Nevertheless we calculated $M/L_{*,i} = 0.17 (M/L)_{\odot}$ using the {\sc pegase.2} models with Kroupa IMF. The formal relative uncertainty of this estimate is the same as for centre, but it is underestimated because doesn't consider the variations of additional parameters utilised for full spectral fitting technique. For comparison we also} used the total magnitude in the $i$ band and the $g-i$ colour from aperture photometry, outside bright SF regions to estimate $\Upsilon(i,g-i) = 0.21 (M/L)_{\odot}$ using the formula from \cite{Zibetti2009}. \revone{Both estimates are consistent within uncertainties.} Then the corresponding stellar masses are $3.2 \times 10^8$ for the inner central component, and \revone{$5.4 \times 10^7$ or $6.7 \times 10^7$ for the LSB disc (obtained using SED fitting results or formula from \cite{Zibetti2009}, respectively). It gives us the stellar mass ratio of components at least $\sim5:1$.}

\subsection{\revone{The mass modelling of the rotation curve}}

\begin{figure*}

\includegraphics[trim=0.5cm 0 1cm 0,clip,height=0.21\textwidth ]{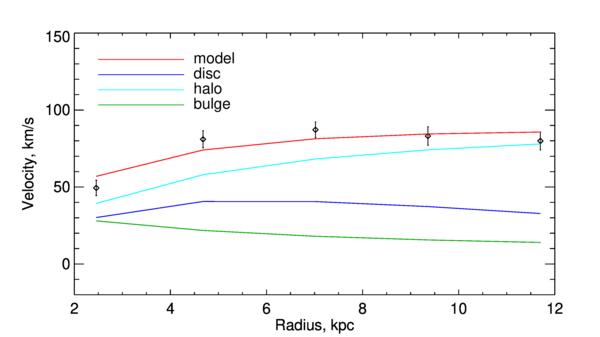}
\includegraphics[trim=2.40cm 0 1cm 0,clip,height=0.21\textwidth]{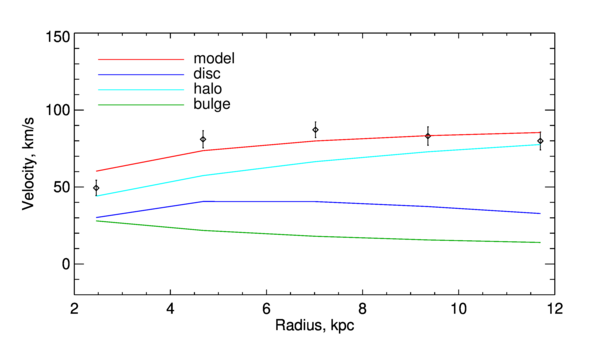}
\includegraphics[trim=2.40cm 0 1cm 0,clip,height=0.21\textwidth]{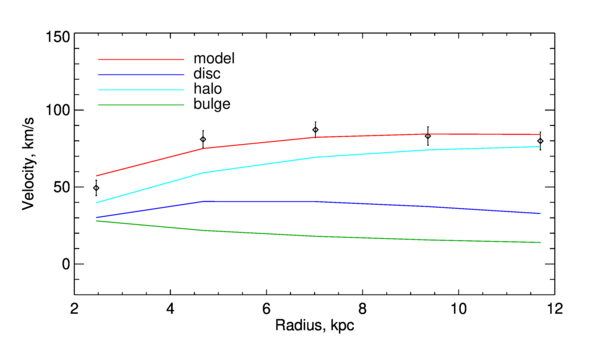}

\includegraphics[height=0.25\textwidth]{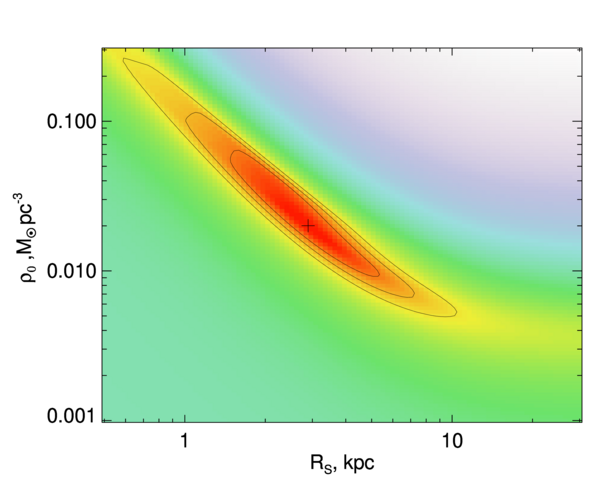}
\includegraphics[height=0.25\textwidth]{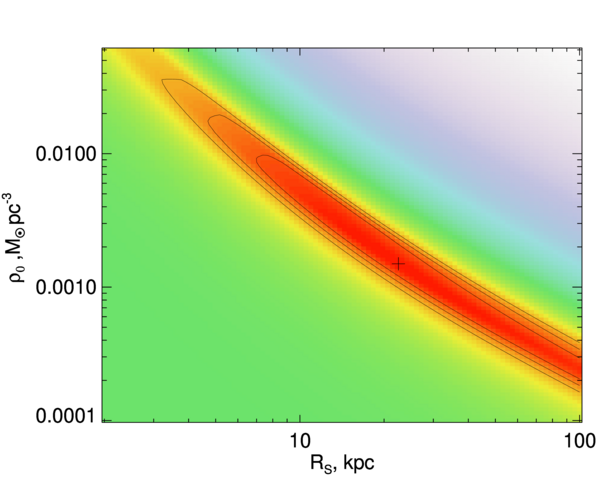}
\includegraphics[height=0.25\textwidth]{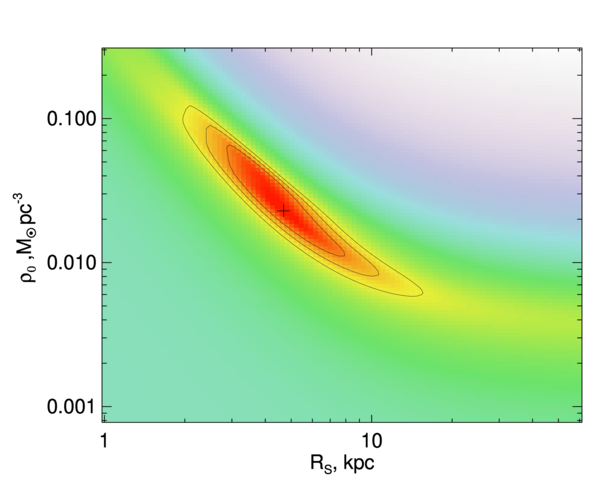}

\caption{Top panel: the best-fitting models of the rotation curve of Ark 18 left ~--- for the pISO profile of the DM halo, centre ~--- for the NFW profile, right ~--- to the Burkert profile.
Bottom panel: $\chi^2$ map for the parameters of dark halo, colour in the maps denotes the $\chi^2$ value, the darker the color, the lower the $\chi^2$ and the better is the fitting quality.
The contours refer to $1\sigma$, $2\sigma$ and $3\sigma$ confidence limits.
The position of the parameters corresponding to the $\chi^2$ minimum is shown by the cross in each map.}
\label{rcmod}
\end{figure*}

\begin{table}
\begin{center}
\caption{The derived parameters of the dark halo with $1\sigma$ confidence limit errors. The columns contain the following data:
(1)~-- dark halo profile;
(2) and (3)~-- radial scale and central density of the DM halo;
(3)~-- mass of DM halo inside of radius four disc radial scales 10.9 kpc \label{par}}
\renewcommand{\arraystretch}{1.5}
\begin{tabular}{lrlrl  rlrlrl}
\hline
dark halo	&	\multicolumn{2}{c}{$R_s$}&	\multicolumn{2}{c}{$\rho_0$ }&		\multicolumn{2}{c}{$M_{\rm halo}$}	 \\
&\multicolumn{2}{c}{kpc}&\multicolumn{2}{c}{$10^{-3}$ M$_{\odot}/$pc$^3$}&\multicolumn{2}{c}{$10^{10}$ M$_{\odot}$}\\

\hline
\hline
          Burkert&        4.62 & $^{+     2.97}_{-     1.56 } $  &       22.86& $^{+     40.44}_{-      11.72} $  &       1.40& $^{+     0.66}_{-    0.28} $    \\                     NFW&    22.28&    &      1.51&    &         1.47&        \\
                   pISO&     2.84& $^{+     2.02}_{-     1.25} $  &      20.00& $^{+      43.32}_{-      10.55} $  &        1.46& $^{+     0.63}_{-     0.28} $    \\\hline
\hline
\end{tabular}
\end{center}
\end{table}

To estimate the contributions of the \revone{central component}, the LSB disc and the dark matter halo to the total mass of Ark~18, we performed the mass modelling of the rotation curve obtained from the FPI data. During the decomposition we used the following components: a Sersic inner component, an exponential disc and a dark matter halo. We utilized three different profiles of dark halo: NFW \citep{nfw}, Burkert \citep{burkert} and a pseudo-isothermal profile.  The details of the method of the rotation curve decomposition are described in \citet{saburova2016}.

For the mass modelling we could use only the part of the rotation curve outside the radius of $R=20$~arcsec, since the data in the central region are influenced by the feedback \revone{(see below in Section~\ref{sec:broadcomp})} and the non-circular motions appear to be comparable with the velocity of rotation. During the modelling we fixed the structural parameters of the inner component and the LSB disc to the values obtained from the {\sc galfit} {\it i}-band image decomposition. The {\it i}-band mass-to-light ratio of the central component was fixed to 0.39 (in Solar units) following the results of the SED+spectrum fitting (see Section~\ref{sec:sed}). The mass-to-light ratio of the LSB disc was set to 6.88 (Solar units). \revone{In the disc mass-to-light ratio, the mass includes both stellar mass calculated from the disc {\it i}-band luminosity and $M/L_i$ estimated from the colour index in Section \ref{sec:sed} and the gas mass from Table \ref{tab:summary} taking into account the contribution of Helium.  The gas mass} is significantly higher than the stellar mass. The adopted masses of the baryonic components are as follows: $3.2\times10^{8}$ M$_{\odot}$  and $2.8\times10^{9}$ M$_{\odot}$ for the central component and LSB+\HI disc correspondingly. We present the result of mass modelling in Fig.~\ref{rcmod} for the pseudo-isothermal (left-hand panel), NFW (middle panel) and Burkert (right-hand panel) dark halo profiles. The top panel shows the model of the rotation curve and bottom panel demonstrates $\chi^2$ maps for the central density and the radial scale of the dark halo. The contours refer to $1\sigma$, $2\sigma$ and $3\sigma$ confidence levels. It is evident from Fig.~\ref{rcmod} that the NFW profile poorly describes the observed rotation curve because we obtain the infinite $1\sigma$ contour on the $\chi^2$ map.

We present the derived parameters of the dark halo in Table~\ref{par}, from which one can see that Ark~18's dark matter halo is not extreme. Its radial scale and central density lie within the range typically found in normal discy galaxies for given disc radii (see e.g. \citealt{saburovas2018} for comparison). It could indicate that the unusual appearance of Ark~18 is not related to the unusual properties of its dark halo, instead it should be explained by some other reasons, e.g. the history of the baryon mass assembly. 

\subsection{Gas excitation, chemical abundance and stellar feedback}\label{sec:emission}

\begin{figure}
    \centering
    \includegraphics[width=\linewidth]{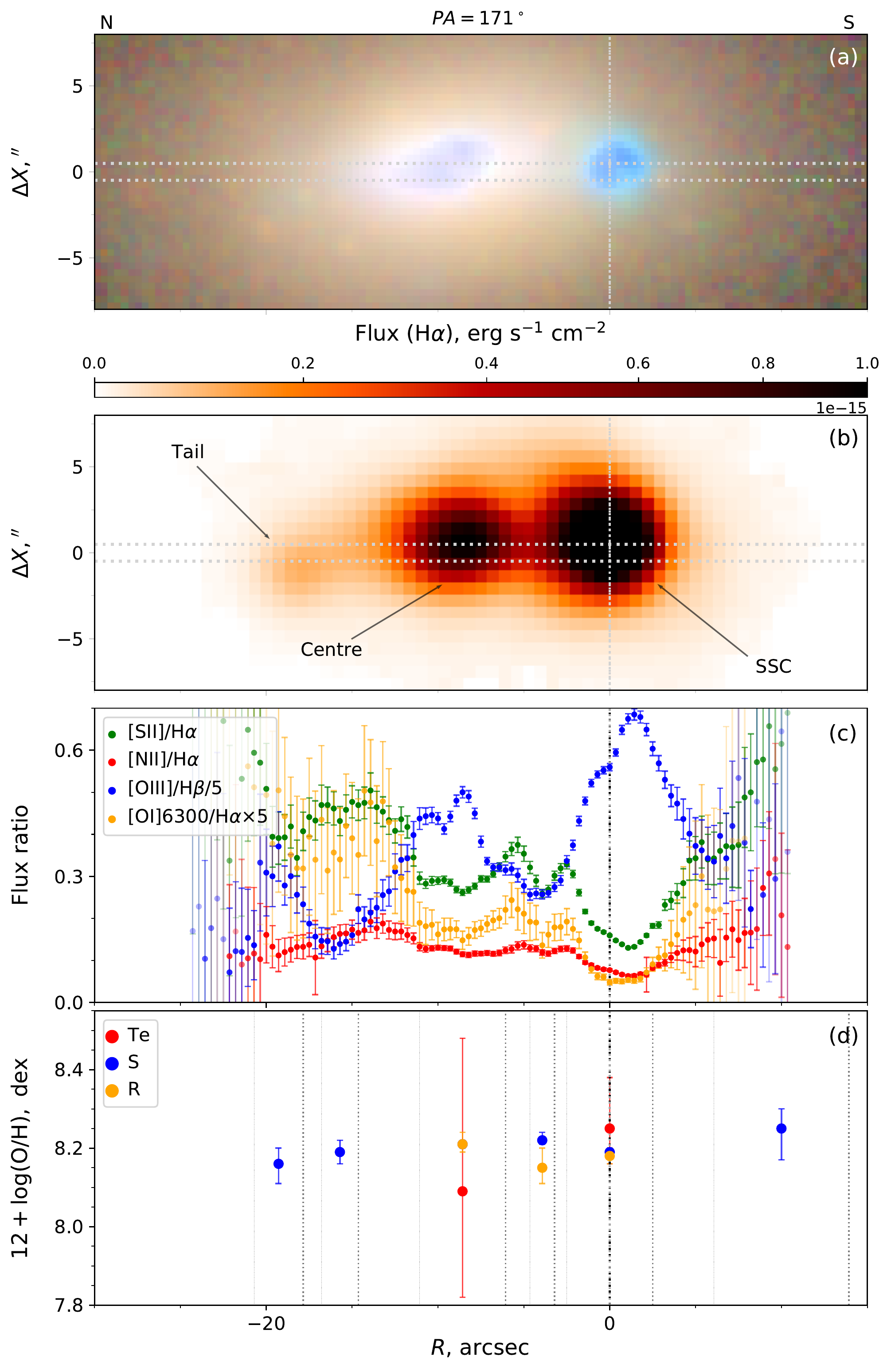}
    \caption{\revone{Results of the analysis of the emission-line spectrum obtained at PA=171. The position of the slit is overlaid on (a) SDSS $gri$ colour image and (b) \Ha flux map by horizontal dotted lines. Panels (c) and (d) represent the distribution along the slit of the flux ratios and oxygen abundance, respectively. The oxygen abundance derived with three methods ($T_e$, S and R described in Section~\ref{sec:abundances}) are shown by different colours.} The vertical lines on panel (d) show the limits of each region where the spectrum was integrated to estimate the oxygen abundance.}
    \label{fig:lsresults_pa171}
\end{figure}

\begin{figure*}
    \centering
    \includegraphics[width=\linewidth]{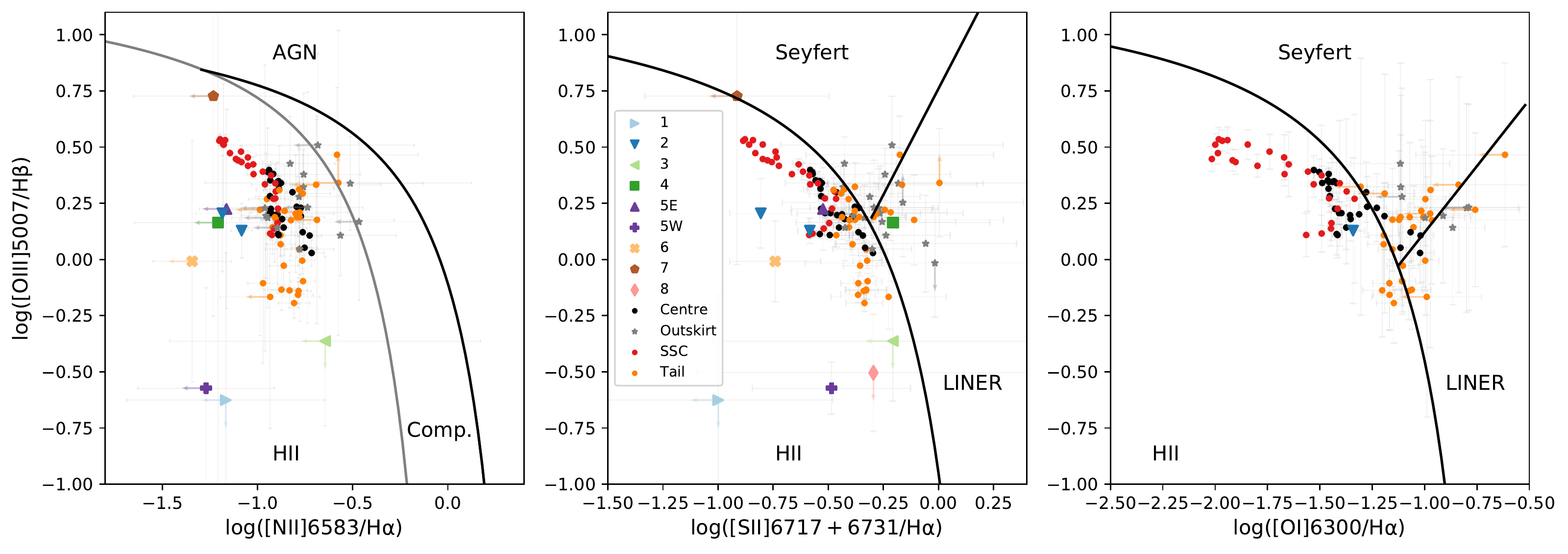}
    \includegraphics[width=\linewidth]{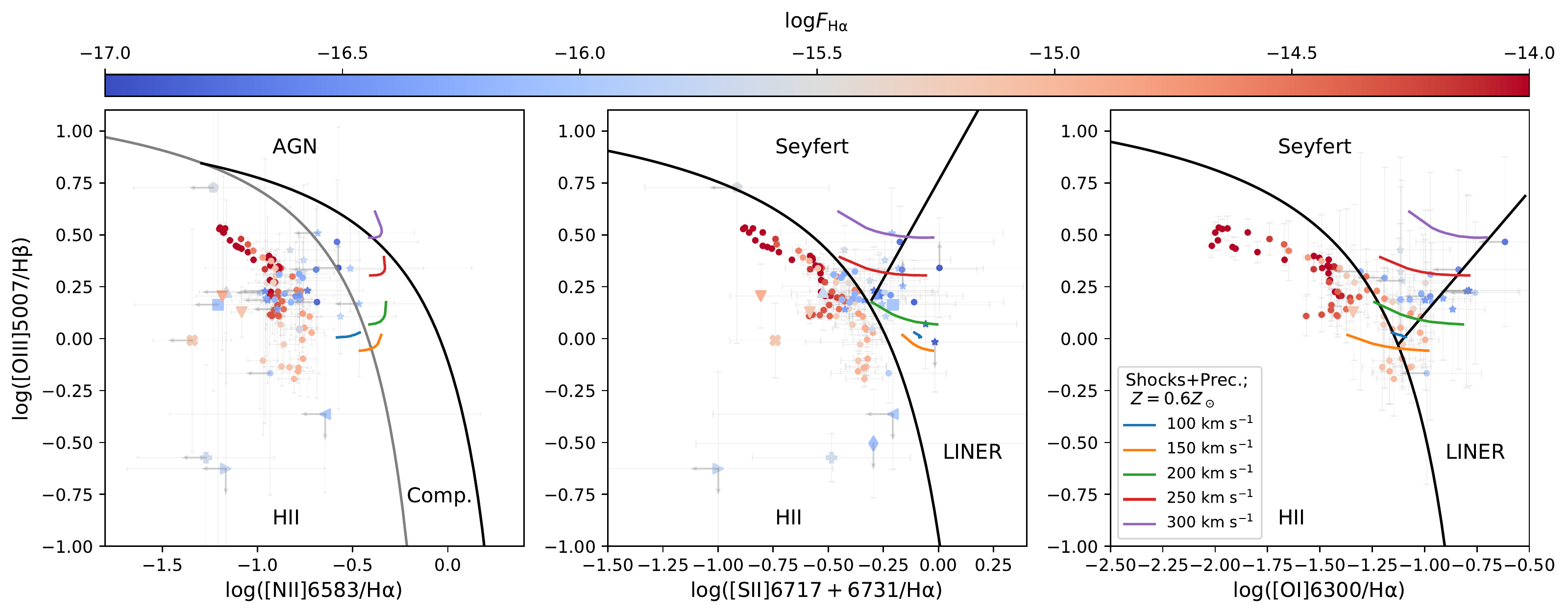}
    \includegraphics[width=\linewidth]{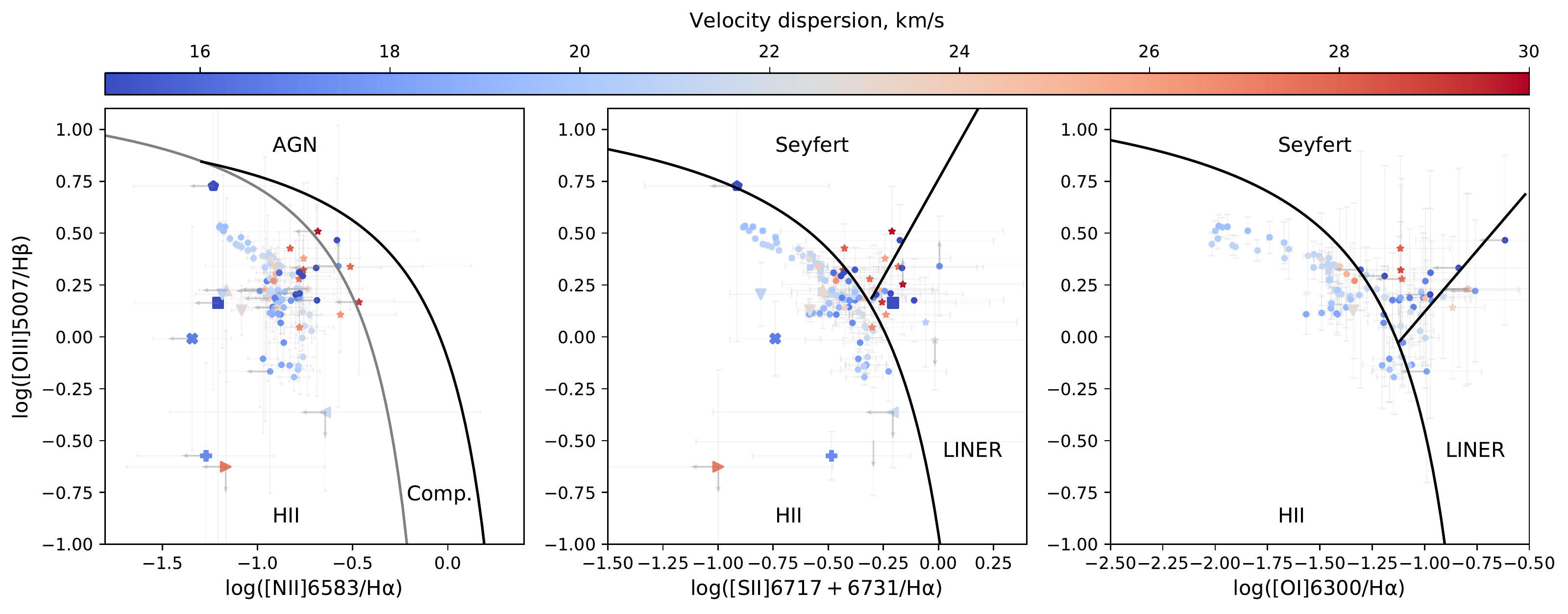}
    \caption{Diagnostic BPT diagrams for \revone{\HII regions in Ark~18 colour coded by the region (top row; names are given according to Fig.~\ref{fig:slitpos} and Tables~\ref{tab:spectral_par_bulge},~\ref{tab:spectral_par_lsb}), its \Ha surface brightness (middle row) and \Ha velocity dispersion (bottom row).  Circles and asterisks show the values for the SSC, ``tail'' and ``outskirt'' obtained for each pixel along the slit within the corresponding region, while the values shown by other symbols were obtained from integrated spectra of a respective region.} Black and grey curved lines separating the areas of different mechanism of excitation are from \citet{Kewley2001} and \citet{Kauffmann2003}, respectively, while the straight line is from \citet{Kewley2006}. Colour lines in the middle panels are the lines of constant velocity of modelled shocks at $Z=0.6Z_\odot$ according to \citet{Allen2008}.}
    \label{fig:bpt}
\end{figure*}

The chemical abundance of the interstellar medium in Ark~18 has never been studied in detail. The only available estimate of the oxygen abundance, which is the most widely-used indicator of the gas-phase metallicity, was obtained by \cite{Eridanus}. They provide the value of $\mathrm{12+\log(O/H)}=8.08\pm0.12$~dex derived from the SDSS spectrum of the SSC region only. The authors also measured the oxygen abundance in the centre of the galaxy (based on the SDSS spectrum) and obtained a significantly lower value of $\mathrm{12+\log(O/H)}\sim7.55$~dex yet \revone{due to the limitations of this estimate (that we discuss in Sec.~\ref{sec:abundances}) authors didn't rely on it in their analysis.} 
Such large discrepancy between the metallicity at the centre of the galaxy and the SSC clump (indicative of the inverted metallicity gradient) motivated us to perform more detailed analysis of the oxygen abundance in Ark~18 based on the long-slit spectra.

In Fig.~\ref{fig:slitpos} we show the slit positions during our spectroscopic observations on top of the SDSS and \Ha images, and also give the designation of the crossed regions that will be used in the subsequent analysis. The slit position $PA=171^\circ$ crosses the SSC and centre of the galaxy, while the slits $PA=118^\circ$ and $PA=145^\circ$ provide the information for 8 star-forming clumps in the LSB disc (hereafter referred to by the numbers\footnote{We considered Western and Eastern parts of the region \#5 separately as 5W and 5E because their fluxes in \OIII line drastically differ.}) and also for periphery of the central \revone{component} (hereafter referred to as ``the outskirt''). In particular, the slits $PA=171^\circ$ and $PA=118^\circ$ cross the faint ``tail'' of ionised gas visible in the \Ha distribution on the Northern part of the galaxy's \revone{central component} -- we were mentioned it before in Sec.~\ref{sec:photometry} and will consider it in a greater detail below. From the reduced spectra we were able to extract the spatial distribution of properties along the slits only for the central part of the galaxy (including the ``tail''). In Fig.~\ref{fig:lsresults_pa171} we show the distribution of the derived parameters (flux ratios and oxygen abundance) along the slit $PA=171^\circ$, while for the regions in the LSB disc we give the estimates made by its integrated co-added spectrum. The de-reddened fluxes of emission lines measured from integrated spectra for \HII regions in the \revone{central component} and in the LSB disc are given in Tables~\ref{tab:spectral_par_bulge} and \ref{tab:spectral_par_lsb}, respectively.

\subsubsection{Gas excitation}

Fig.~\ref{fig:lsresults_pa171} clearly shows significant variations of the flux ratios of main diagnostic emission lines across the \revone{central component} of Ark~18. In particular, the ratio of \OIIIHb\ grows up towards the centre and especially towards the SSC where it reaches the maximum value in the galaxy (\OIIIHb$\sim3.5$). Note also that both ``the centre'' and SSC regions demonstrate double peaks in the \OIIIHb\ distributions suggesting that probably at least two main sources are responsible for ionisation of each of the two brightest \HII regions. In fact, the two star clusters in the SSC region are resolved in archival VISTA VHS images. The intensity of the low excitation lines is enhanced in the outskirts and towards ``the tail''. The ratio of \SIIHa>0.4 is typical for diffuse ionised gas (DIG), an extended component of ionised gas of low excitation that could account for up-to 50 per cent of the total \Ha luminosity of a galaxy \citep{Oey2007}. Among possible sources of the DIG ionisation are the leaking ionising photons filtered by \HII regions, the shocks in a low density environment, the ionisation by hot low mass evolved stars \citep[see, e.g.,][for review]{Haffner2009,Zhang2017}. As we show later, in Ark~18 several of these mechanisms might be responsible for the DIG excitation. 

In Fig.~\ref{fig:bpt} we plot all the studied regions on the widely used BPT diagnostic diagrams (introduced by \citealt*{BPT} and extended by \citealt{Veilleux1987}). In the top row, the diagrams of different colours and symbols denote the different regions according to Fig.~\ref{fig:slitpos} and Tables~\ref{tab:spectral_par_bulge}, \ref{tab:spectral_par_lsb}. For the regions within the \revone{central component} (SSC, ``centre'', ``tail'' and ``outskirt'') we give the values measured from individual pixels along the slits. Every studied region except \#4 from the LSB disc lies below the `maximum starburst line' from \cite{Kewley2001} and even below the \cite{Kauffmann2003} line bounding the area where pure photoionisation by massive stars could be responsible for the gas excitation. The same is true for the SSC and ``centre'', while a part of the ``tail'' and most of the ``outskirt'' fall to the area where different excitation mechanisms (e.g. AGN, shocks, ionisation by low mass evolved stars) should take place. This is especially evident from the [S~\textsc{ii}]/H$\alpha$ and [O~\textsc{i}]/H$\alpha$ diagnostics which are more sensitive to shocks than \NIIHa. 
 
In the middle row of Fig.~\ref{fig:bpt} the surface brightness is shown by colour, while the symbols are the same as in the top row. From these diagrams it is evident that all regions lying above \cite{Kewley2001} and \cite{Kauffmann2003} lines on \NIIHa\ and \SIIHa\ diagnostics have low surface brightness and rather correspond to the DIG emission (following the limiting brightness $L(\mathrm{H\alpha)<10^{39}\ erg\ s^{-1}\ cm^{-2}\ kpc^{-2}}$ from \citealt{Zhang2017}). Note that the models of low-velocity shocks for the metallicity $Z=0.6Z\odot$ could well explain these DIG regions. However, in the bottom row showing how the velocity dispersion in \Ha line varies accross the BPT diagram (a so-called BPT-$\sigma$ diagram, see \citealt{Oparin2018}) one may see that the enhanced velocity dispersion is observed only in the ``outskirt'' region, but not in the ``tail''. It is also enhanced towards the region~\#1, which also has a surface brightness in \Ha that is typical for DIG, and as follows from the Fig.~\ref{fig:FPI}, such a high value of $\sigma_{\mathrm{gas}}$ is observed only at the periphery of the region. From this analysis we may conclude that the emission of the outskirts of the \revone{central component} is probably caused by the shocks because of its specific location on the BPT diagram and the enhanced velocity dispersion. DIG in the ``tail'' has rather different origin and is probably excited by leaking photons from nearby \HII regions. The SSC and the ``centre'' as well as most of the regions in the LSB disc are photoionised by massive stars. The region~\#4 is probably located very close to the \revone{central component's} outskirt, so its emission lines flux ratios could be explained by the significant contamination by the surrounding DIG. 

\begin{table*}
\caption{Measured line fluxes (corrected for reddening with the adopted E(B-V)) normalized to the flux of H$\beta$ = 100, and oxygen abundances for \HII regions in the \revone{central component} of Ark~18}\label{tab:spectral_par_bulge}
\begin{footnotesize}
\begin{tabular}{ccccccc}
\hline
Parameter & Tail & Tail & SSC & Centre & Outskirt & Outskirt \\
\hline
Slit & PA=171$^\circ$ & PA=118$^\circ$ & PA=171$^\circ$ & PA=171$^\circ$ & PA=171$^\circ$ & PA=145$^\circ$ \\
Position & -20.7...-17.9 & -6.7...4.7 & -2.5...2.5 & -11.1...-6.1 & 6.1...13.9 & -69.1...-46.3 \\
{[O~\textsc{ii}]} 3727\AA & -- & -- & $273.6\pm13.8$ & $428.9\pm25.2$ & -- & -- \\
{[Ne~\textsc{iii}]} 3869\AA & -- & -- & $18.2\pm2.0$ & $11.6\pm4.9$ & -- & -- \\
H$\gamma$ & $37.4\pm13.1$ & -- & $45.2\pm0.8$ & $42.9\pm2.0$ & $45.0\pm35.2$ & -- \\
{[O~\textsc{iii}]} 4363\AA & -- & -- & $1.7\pm0.4$ & $1.4\pm1.3$ & -- & -- \\
{He~\textsc{i}} 4471\AA & -- & -- & $3.0\pm0.3$ & -- & -- & -- \\
{[Fe~\textsc{iii}]} 4658\AA & -- & -- & $0.8\pm0.3$ & -- & -- & -- \\
{He~\textsc{ii}} 4686\AA & -- & -- & $0.7\pm0.3$ & -- & -- & -- \\
H$\beta$ & $100.0\pm7.2$ & $100.0\pm11.1$ & $100.0\pm0.4$ & $100.0\pm0.7$ & $100.0\pm15.7$ & $100.0\pm73.5$ \\
{[O~\textsc{iii}]} 5007\AA & $136.6\pm6.0$ & $189.5\pm10.1$ & $287.7\pm2.2$ & $210.6\pm1.7$ & $192.8\pm12.7$ & $201.7\pm49.6$ \\
{He~\textsc{i}} 5015\AA & -- & -- & $2.4\pm0.5$ & -- & -- & -- \\
{[N~\textsc{i}]} 5200\AA & -- & -- & $0.9\pm0.2$ & $1.9\pm0.6$ & -- & -- \\
{[N~\textsc{ii}]} 5755\AA & -- & -- & $0.4\pm0.1$ & -- & -- & -- \\
{He~\textsc{i}} 5876\AA & -- & -- & $10.6\pm0.2$ & $9.2\pm0.6$ & -- & -- \\
{[O~\textsc{i}]} 6300\AA & $21.8\pm5.5$ & $36.8\pm11.5$ & $3.8\pm0.2$ & $9.8\pm0.7$ & $17.6\pm11.9$ & -- \\
{[S~\textsc{iii}]} 6312\AA & -- & -- & $1.1\pm0.1$ & $1.6\pm0.7^*$ & -- & -- \\
H$\alpha$ & $280.9\pm6.1$ & $257.9\pm7.1$ & $281.0\pm3.6$ & $281.2\pm11.4$ & $281.2\pm11.4$ & $268.6\pm43.2$ \\
{[N~\textsc{ii}]} 6583\AA & $35.3\pm3.9$ & $44.1\pm6.1$ & $20.9\pm0.4$ & $33.2\pm0.7$ & $47.5\pm7.6$ & $<36.3\pm35.9$ \\
{[S~\textsc{ii}]} 6717\AA & $68.9\pm4.6$ & $81.1\pm7.2$ & $27.0\pm0.2$ & $47.1\pm0.5$ & $76.2\pm8.3$ & $134.5\pm45.9$ \\
{[S~\textsc{ii}]} 6731\AA & $45.4\pm4.3$ & $59.4\pm7.0$ & $19.2\pm0.1$ & $33.8\pm0.4$ & $55.8\pm8.1$ & $35.2\pm37.7$ \\
{[Ar~\textsc{iii}]} 7136\AA & -- & -- & $6.5\pm0.1$ & $4.3\pm0.5$ & -- & -- \\
{[S~\textsc{iii}]} 9069\AA & -- & -- & -- & $16.5\pm3.1^*$ & -- & -- \\
E(B-V), mag & $0.01\pm0.06$ & $0.00\pm0.10$ & $0.07\pm0.01$ & $0.17\pm0.04$ & $0.14\pm0.14$ & $0.00\pm0.65$ \\
EW(H$\alpha$), \AA & $16.4$ & $7.9$ & $224.2$ & $31.8$ & $22.1$ & $8.9$ \\
$\sigma(\mathrm{H\alpha})$, km s$^{-1}$ & $17.1\pm0.2$ & $20.0\pm4.6$ & $20.7\pm0.3$ & $21.2\pm0.2$ & $29.8\pm2.4$ & $14.5\pm5.5$ \\
$n_e$, cm$^{-3}$ & $-$ & $58_{-16}^{+213}$ & $28_{-9}^{+13}$ & $37_{-18}^{+16}$ & $58_{-18}^{+299}$ & $-$ \\
$T_e$([O~\textsc{iii}]), K & -- & -- & $9663_{-615}^{+653}$ & $12293_{-3190}^{+3936^*}$ & -- & -- \\
$\mathrm{12+\log(O/H)_{Te}}$ & -- & -- & $8.25_{-0.09}^{+0.13}$ & $8.09_{-0.27}^{+0.39}$ & -- & -- \\
$\mathrm{12+\log(O/H)_{S}}$ & $8.16_{-0.05}^{+0.04}$ & $8.22_{-0.07}^{+0.05}$ & $8.19_{-0.01}^{+0.01}$ & $8.21_{-0.01}^{+0.01}$ & $8.25_{-0.08}^{+0.05}$ & $8.14_{-0.24}^{+0.12}$ \\
$\mathrm{12+\log(O/H)_{R}}$ & -- & -- & $8.18_{-0.02}^{+0.02}$ & $8.21_{-0.02}^{+0.02}$ & -- & -- \\
$\mathrm{\log(N/O)_{R}}$ & -- & -- & $-1.33_{-0.03}^{+0.04}$ & $-1.32_{-0.04}^{+0.05}$ & -- & -- \\
$\mathrm{\log(N/O)_{Te}}$ & -- & -- & $-1.24_{-0.01}^{+0.24}$ & $-0.76_{-0.31}^{+0.1}$ & -- & -- \\
$\mathrm{\log(S/O)_{Te}}$ & -- & -- & $-1.64_{-0.07}^{+0.23}$ & $-1.44_{-0.07}^{+0.23}$ & -- & -- \\
$\mathrm{\log(Ne/O)_{Te}}$ & -- & -- & $-0.75_{-0.02}^{+0.29}$ & $-0.36_{-0.44}^{+0.32}$ & -- & -- \\
$\mathrm{\log(Ar/O)_{Te}}$ & -- & -- & $-2.46_{-0.17}^{+0.07}$ & $-2.45_{-0.13}^{+0.15}$ & -- & -- \\
\hline
\multicolumn{7}{l}{$^*$The value is derived from the SDSS spectrum for the entire `centre' region}\\
\end{tabular}
\end{footnotesize}
\end{table*}

\begin{table*}
\caption{\revone{Same as Table~\ref{tab:spectral_par_bulge} but for regions in LSB disc}}\label{tab:spectral_par_lsb}
\begin{footnotesize}
\begin{tabular}{cccccc}
\hline
Parameter & 1 & 2 & 2 & 3 & 4  \\
\hline
Slit & PA=118$^\circ$ & PA=145$^\circ$ & PA=118$^\circ$ & PA=118$^\circ$ & PA=118$^\circ$ \\
Position & -78.1...-67.4 & -114.1...-107.0 & -51.7...-44.6 & -33.5...-20.6 & 12.2...20.1   \\
H$\beta$ & $100.0\pm37.4$ & $100.0\pm29.7$ & $100.0\pm10.5$ & $100.0\pm88.0$ & $100.0\pm81.6$ \\
{[O~\textsc{iii}]} 5007\AA & $<23.7$ & $160.2\pm19.9$ & $134.2\pm9.2$ & $<43.3$ & $145.8\pm86.7$   \\
{[O~\textsc{i}]} 6300\AA & -- & -- & $12.2\pm9.1$ & -- & -- \\
H$\alpha$ & $190.1\pm24.7$ & $281.6\pm15.4$ & $268.4\pm7.1$ & $236.5\pm51.8$ & $273.6\pm68.6$  \\
{[N~\textsc{ii}]} 6583\AA & $<12.9$ & $18.4\pm10.9$ & $22.1\pm6.1$ & $<53.6$ & $<17.0$  \\
{[S~\textsc{ii}]} 6717\AA & $7.1\pm12.3$ & $31.5\pm11.5$ & $38.8\pm6.1$ & $54.8\pm50.8$ & $88.8\pm53.9$  \\
{[S~\textsc{ii}]} 6731\AA & $12.0\pm16.2$ & $12.5\pm10.2$ & $31.2\pm6.4$ & $92.1\pm67.4$ & $81.3\pm55.5$  \\
E(B-V), mag & $0.00\pm0.34$ & $0.36\pm0.26$ & $0.00\pm0.09$ & $0.00\pm0.78$ & $0.00\pm0.74$  \\
EW(H$\alpha$), \AA & $197.5$ & $361.0$ & $299.5$ & $28.0$ & $8.1$ \\
$\sigma(\mathrm{H\alpha})$, km s$^{-1}$ & $27.3\pm8.7$ & $20.9\pm1.9$ & $22.8\pm6.9$ & $21.6\pm6.9$ & $13.0\pm4.7$ \\
$\mathrm{12+\log(O/H)_{S}}$ & $<8.45$ & $8.28_{-0.54}^{+0.06}$ & $8.15_{-0.33}^{+0.07}$ & $8.37_{-0.25}^{+0.12}$ & $8.1_{-0.29}^{+0.14}$ \\
\hline
\hline
Parameter & 5E & 5W & 6 & 7 & 8 \\
\hline
Slit & PA=118$^\circ$ & PA=118$^\circ$ & PA=145$^\circ$ & PA=145$^\circ$ & PA=118$^\circ$ \\
Position & 47.6...53.3 & 41.8...46.1 & -6.3...3.7 & 6.6...12.3 & 26.8...34.0 \\
H$\beta$ & $100.0\pm29.0$ & $100.0\pm27.4$ & $100.0\pm38.9$ & $100.0\pm89.0$ & $100.0\pm68.1$ \\
{[O~\textsc{iii}]} 5007\AA & $167.8\pm26.6$ & $26.7\pm20.8$ & $98.1\pm26.7$ & $533.3\pm84.8$ & $<31.3$ \\
H$\alpha$ & $232.7\pm18.5$ & $272.3\pm17.0$ & $281.3\pm18.9$ & $281.6\pm52.0$ & $197.8\pm52.2$ \\
{[N~\textsc{ii}]} 6583\AA & $<16.0$ & $<14.6$ & $<12.8$ & $<16.5$ & -- \\
{[S~\textsc{ii}]} 6717\AA &  $43.1\pm21.0$ & $51.3\pm16.7$ & $30.6\pm15.2$ & $<23.1$ & $60.0\pm38.9$ \\
{[S~\textsc{ii}]} 6731\AA & $26.5\pm20.0$ & $37.6\pm16.1$ & $20.5\pm14.4$ & $<11.2$ & $40.3\pm37.6$ \\
E(B-V), mag & $0.00\pm0.26$ & $0.00\pm0.24$ & $0.24\pm0.34$ & $0.38\pm0.84$ & $0.00\pm0.63$ \\
EW(H$\alpha$), \AA & $169.2$ & $438.1$ & $227.7$ & $274.8$ & $9.5$  \\
$\sigma(\mathrm{H\alpha})$, km s$^{-1}$ & $23.0\pm4.4$ & $17.3\pm1.4$ & $16.7\pm3.6$ & $10.6\pm3.6$ & $-$ \\
$\mathrm{12+\log(O/H)_{S}}$ & $8.1_{-0.39}^{+0.08}$ & $8.04_{-1.01}^{+0.02}$ & $8.19_{-0.79}^{+0.03}$ & $8.35_{-0.24}^{+0.1}$ & -- \\
\hline
\end{tabular}
\end{footnotesize}
\end{table*}

 \subsubsection{Gas phase chemical abundances}
 \label{sec:abundances}
 
To precisely estimate the gas chemical abundance, it is necessary to measure at least one of the faint lines sensitive to the electron temperature ($T_e$): [O~\textsc{iii}]~4363\AA, [N~\textsc{ii}]~5755\AA, or [S~\textsc{iii}]~6312\AA\ (together with [S~\textsc{iii}]~9069\AA). Knowing the electron temperature, one can estimate the abundance of different ions using a so-called $T_e$ method. In the case of Ark~18 we were able to measure the faint auroral lines only in the SSC and ``centre'' regions. For the SSC we detected [O~\textsc{iii}]~4363\AA, [N~\textsc{ii}]~5755\AA\ lines and derived $T_e$ in both low and high excitation zones independently. For the ``centre'' region we detected only [O~\textsc{iii}]~4363\AA\ line, and its flux was very uncertain. We analysed also the available SDSS spectrum and found that the flux of [O~\textsc{iii}]~4363\AA\ line there differs significantly while the S/N is still very low (about 2). Because of that, we decided not to use this line to estimate the electron temperature of the ``centre'' region, but instead measured $T_e$ from the flux ratio of [S~\textsc{iii}]~9069\AA\ to [S~\textsc{iii}]~6312\AA\ lines (both are available in the SDSS spectrum). The obtained value of $T_e$ corresponds to the areas of high excitation for S$^{2+}$ and Ar$^{2+}$ ions, but could be converted to $T_e$ for other ions following the relations from \citet{Garnett1992}:
 \begin{equation}
T_e(\mathrm{[S~\textsc{iii}]})=0.83T_e(\mathrm{[O~\textsc{iii}]})+1700,
 \end{equation} \begin{equation}
 T_e(\mathrm{[N~\textsc{ii}]})=0.7T_e(\mathrm{[O~\textsc{iii}]})+3000
 \end{equation}
 
At the end, we derived the value of $T_e$ from our observations for SSC and estimated it from the SDSS spectrum for the ``centre'' region. The electron density $n_e$ was estimated from the ratio of [S~\textsc{ii}] 6717~\AA\ and 6731~\AA\ lines. Using these physical parameters, we derived the oxygen abundance $\mathrm{12+\log(O/H)}$ as well as the relative abundances of N/O, S/O, Ar/O, and Ne/O with the \textsc{PyNeb} package \citep{Luridiana2015}. To correct the measured ionic abundances for N, S, Ar, Ne for unobserved species we used the ionisation correction factors from \citet{Izotov2006}. The final abundances are given in Table~\ref{tab:spectral_par_bulge}, and the oxygen abundance is also shown in Fig.~\ref{fig:lsresults_pa171}.
 
To estimate the oxygen abundance in all studied regions where only strong emission lines were measured we applied R and S empirical calibrations from \cite{Pilyugin2016}. Both these methods are valid in our range of metallicities 
and provide the values that well agree with the estimates made by the $T_e$ method. The R method is based on the relation between the flux ratios of [O~\textsc{ii}], [N~\textsc{ii}] and [O~\textsc{iii}] lines to H$\beta$, while the S method utilizes [S~\textsc{ii}]/H$\beta$ instead of [O~\textsc{ii}]/H$\beta$. Because of the poor quantum efficiency in the long-slit mode at $\lambda<4000$~\AA, we were able to apply the R method only to the bright central regions. In order to get rid of the uncertainties with the interstellar extinction, we slightly modified R and S methods and normalized the fluxes of \NII and \SII lines to \Ha assuming the theoretical ratio of H$\alpha$/H$\beta=2.86$ for $T_e=10000$~K \citep{Osterbrock2006}. Such modification does not change the results, but reduces the final uncertainty. 
 
\begin{figure*}
    \centering
    \includegraphics[width=0.89\linewidth]{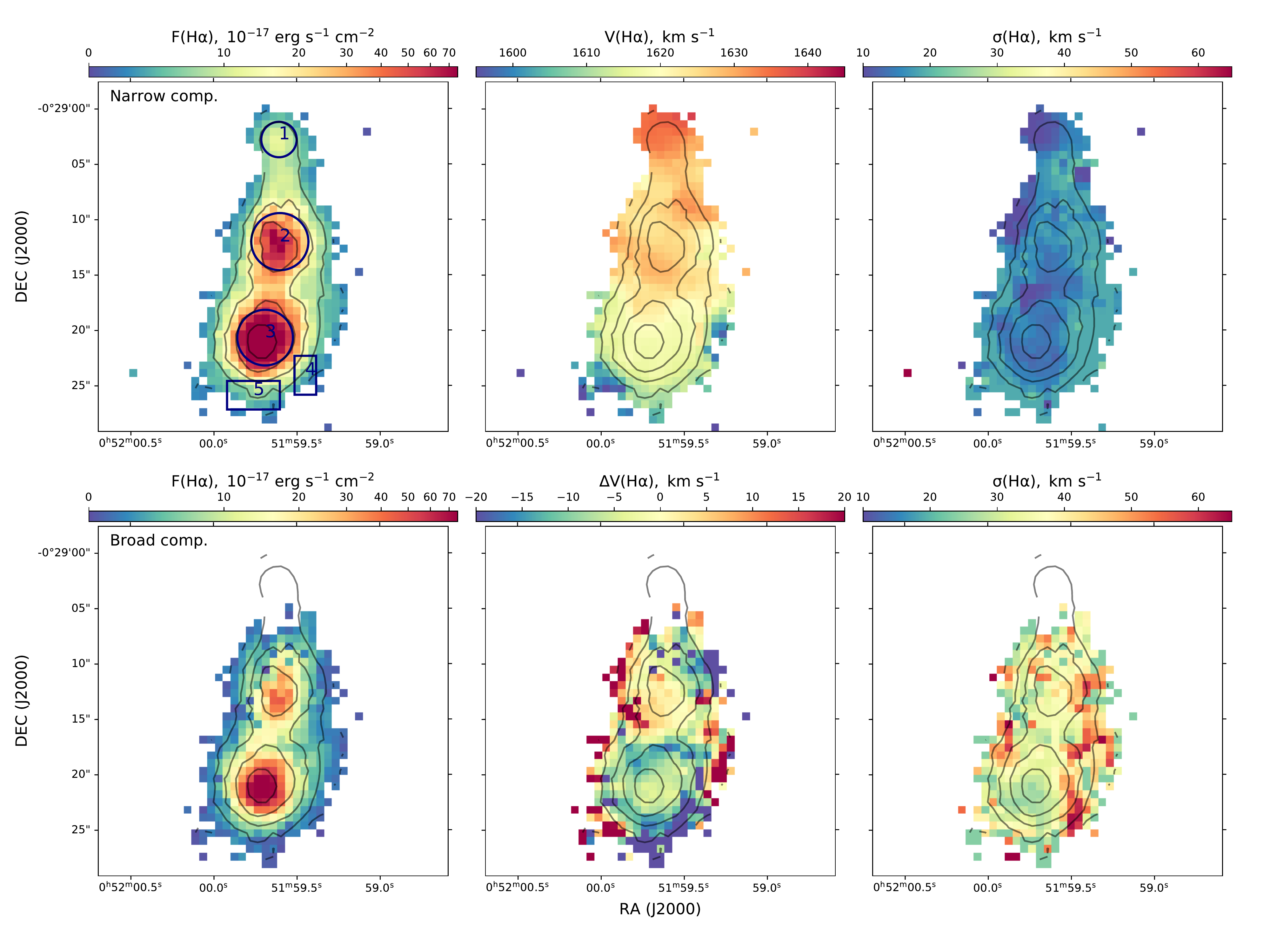}
    \includegraphics[width=\linewidth]{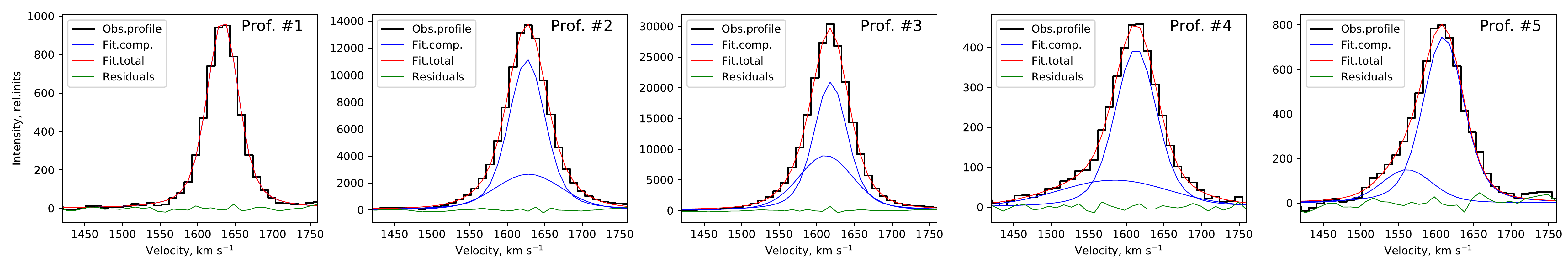}
    \caption{Decomposition of the \Ha data cube into narrow (top row) and broad (middle row) components for the central part of the galaxy. All pixels with S/N$<$15 were masked. From left to right: the distribution of the \Ha flux, line-of-sight velocity (for the broad component its deviation $\Delta V(\mathrm{H\alpha}$) from that of narrow component)  and velocity dispersion for corresponding component. The overlaid contours correspond to the distribution of \Ha flux for the narrow component. The bottom row demonstrates the examples of observed line profiles (black line) integrated over the regions shown in top-left panel, their decomposition onto individual components (blue line), resulting model (red line) and residuals (green line).}
    \label{fig:broadcomp}
\end{figure*}

To estimate the uncertainty of the adopted methods we applied them to Monte-Carlo simulated synthetic observations with the line fluxes randomly distributed around the measured ones with a standard deviation of the probability distributions equal to the measured uncertainties of the fluxes. For several regions in the LSB disc the S/N of \NII \revone{(and in some cases also of [O~\textsc{iii}]) line} was insufficient for its certain measurements and its fluxes appear rather as an upper limit -- in such cases we adopted the probability distribution of its fluxes as uniform from zero to $2\sigma$, where $\sigma$ is the noise level. Our analysis of the probability distribution of the results showed that the value of $\mathrm{12+\log(O/H)}$ obtained from the measured fluxes well agrees with the most probable value for every region except a few. The exceptions are the regions \#4, 5W, 6 and also 2 for the slit PA=145$^\circ$. For these regions the probability distribution of the resulting $\mathrm{12+\log(O/H)_S}$ is bi-modal, and the formally measured values correspond to its secondary minimum which gives significantly ($0.3-0.9$~dex) lower oxygen abundance than the most probable value. For the first three regions such a behaviour is probably related to very uncertain measurements of the \NII line. Because of this, we adopt as a final estimates the values of $\mathrm{12+\log(O/H)}$ at the maximum of the probability distribution instead of those derived from the observed fluxes. 
 
Finally, we derived the metallicity for 7 regions in the LSB disc, 3 regions in the central \revone{component} and also for its outskirt. Surprisingly, we do not see any metallicity gradient in the galaxy. The metallicity varies insignificantly (given the large extent of the LSB disc) -- according to the estimates made with the S method, $\mathrm{12+\log(O/H)}=8.1-8.3$~dex. We do not observe an inverse metallicity gradient mentioned above. Our estimates of $\mathrm{12+\log(O/H)}$ made with $T_e$ and the other two methods for two bright regions agree well (however, less significant for the ``centre'' as mentioned above). The oxygen abundance for the SSC is slightly higher than that by \cite{Eridanus}, but still agrees within uncertainties. 
\revone{There is also a difference in abundances of central region compared to \cite{Eridanus} who used the methods designed for lower metallicities than what we measured. Also because of the limited spectral range of SDSS data, they recovered the flux of [O~\textsc{ii}]~3727 \AA\, from [O~\textsc{ii}]~7320,7330 \AA, that may lead to additional uncertainties. In our analysis we are free of these limitations.}
 
The relative abundances of other elements (N/O, S/O, Ne/O, Ar/O) for the SSC region agree well with the general trends with metallicity as observed in the other galaxies \citep[see, e.g.,][]{Izotov2006} that points to the normal chemical evolution of the region in the recent past. This is not true for the ``centre'' region that shows significantly larger values of these ratios (except for argon) derived with the $T_e$ method. In general, this might be interpreted as a sign of a recent metal-poor gas accretion, however in our case the uncertainties are very high and this conclusion is rather not reliable. However, even within the uncertainties the measured value of (N/O)$_{Te}$ is still higher than that for the SSC and than what is expected for the derived metallicity. This effect could be related to the composite nature of the region as it follows from flux ratios distribution (Fig.~\ref{fig:lsresults_pa171}) and from the morphology visible in SDSS and VISTA VHS images. As it was shown by \cite{Pilyugin2012}, the $T_e$ method could slightly underestimate the metallicity that mimics the enhancement the of N/O ratio, while this effect is not observed for empirical methods. Because our estimates made by the S method are in agreement, we suppose that it might be the case for ``centre'' region and believe that this method produces a more reliable estimate.
 
Summing up, we may conclude that Ark~18 has a flat metallicity gradient with the central oxygen abundance $\mathrm{12+\log(O/H)}=8.20\pm0.04$~dex derived with the S method and consistent with the $T_e$ method.
Note, however, that we cannot completely rule out a possibility of the presence of a low-metallicity gas in the LSB disc. At least our ``formal'' measurements for several regions made with the S method contradicts to the most probable values (adopted as a final estimate) and points to a much lower metallicity. Deeper spectra of the LSB disc providing good enough S/N for \NII or [O~\textsc{ii}] emission lines are necessary to resolve this uncertainty. 

\subsubsection{Local kinematics of ionised gas in star-forming regions}\label{sec:broadcomp}

While in general the \Ha emission line profile  in Ark~18 is well described by a single Voigt component, we do observe underlying broad component in the central part of the galaxy. In the bottom panel of Fig.~\ref{fig:broadcomp} we show the examples of \Ha line profiles integrated over the ``tail'', the ``centre'', the ``SSC'' and the ``outskirt'' of the \revone{central component} to the South-West and South from the SSC. As we mentioned before, a well pronounced minimum of the velocity dispersion is observed in the bright part of the `tail' region, and $\sigma_{\mathrm{gas}}$ is significantly higher in the ``centre'' and SSC regions (see Fig.~\ref{fig:FPI}).
Indeed, the profile \#1 (``tail'') could be easily described by a single narrow Voigt profile with $\sigma_{\mathrm{gas}}=13\ \kms$ (after correction for thermal and natural broadening), and remaining profiles are much better described by two-component profiles with one narrow ($\sigma_{\mathrm{gas}}=14-19\ \kms$) and one broad ($\sigma_{\mathrm{gas}}=28-69\ \kms$) components. A single Voigt profile with $\sigma_{\mathrm{gas}} \simeq 22-24\ \kms$ 
leads to much larger residuals significantly exceeding the noise level.

We performed a decomposition of the \Ha line profile into two Voigt components in each pixel for the \revone{central component} with S/N$>$15 and where the residuals after the subtraction of a single-component model was greater than 3$\sigma$. Almost each pixel fitted by a two-component model revealed a broad underlying component. As a result we obtained separate maps of \Ha fluxes, line-of-sight velocities and velocity dispersions for narrow and broad components (from left to right in the top and middle rows of Fig.~\ref{fig:broadcomp}). After this, we found that the velocity dispersion of a narrow component is similar to that in the ``tail'' and the regions of star formation in the LSB disc, contrary to Fig.~\ref{fig:FPI} where single-component fitting was applied. Thus, such broadening is related to the underlying component rather than to the generally enhanced velocity dispersion. Adding of these second components to the fitting procedure does not change the line-of-sight velocity distribution, so the analysis presented in Section~\ref{sec:struct_and_kin} remains reliable. 

From the obtained maps of the \Ha flux distribution for narrow and broad components, we estimated the relative contribution of the broad component to the total \Ha flux for both the SSC and the ``centre'' as 36 and 32 per cent respectively. Then we estimated the fluxes in the \Ha line from broad component as measured in circular apertures: $F\mathrm{(H\alpha)=3.3\times10^{-14}\ erg\ s^{-1}\ cm^{-2}}$ for the SSC and $\mathrm{1.5\times10^{-14}\ erg\ s^{-1}\ cm^{-2}}$ for the ``centre''. From the fitting of the profiles \#2 and \#3 in Fig.~\ref{fig:broadcomp} we estimated the mean velocity dispersion of the broad component: $\sigma_{\mathrm{gas}}=30\ \kms$ and $39\ \kms$ for the SSC and the ``centre'', respectively. Being not very high, such values are still significantly larger than the velocity dispersion of a narrow component and indicate the presence of supersonic motions in these regions. 

The intensity of both broad and narrow components grow towards the centre of \HII regions that allows us to suggest that supernovae and stellar winds are the most plausible sources of the broad underlying component, which in fact should be a feedback-driven outflow from there.
In particular, it is expected that among other feedback sources, supernovae contribute most to the observed turbulence of the ISM in galaxies \citep{Bacchini2020}. We can check if the supernovae alone are able to produce the outflowing turbulent components with the properties we derived. We can roughly estimate the mass $M_\mathrm{H\textsc{ii}}$ of the outflowing ionised gas in each region assuming that it represents a Str{\"o}mgren sphere:

\begin{equation}
M_\mathrm{H\textsc{ii}} = \frac{4}{3}\pi R_S^3 n_e m_p,
\end{equation}
where $n_e$ is electron density, $m_p$ is a mass of proton and $R_S$ is a radius, that could be calculated as

\begin{equation}
R_S = \left(\frac{3Q_0}{4\pi\alpha_B n_e^2} \right)^{1/3} = \left(\frac{3L(\mathrm{H\alpha})}{4\pi h \nu_\mathrm{H\alpha} \alpha_\mathrm{H\alpha}^\mathrm{eff} n_e^2} \right)^{1/3},
\end{equation}
where $Q_0$ is a number of ionising photons with $\lambda<912$~\AA, $L(\mathrm{H\alpha})$ is a luminosity in the \Ha line, $\alpha_B$ and $\alpha_\mathrm{H\alpha}^\mathrm{eff}$ are the total recombination coefficient of hydrogen and the \Ha effective recombination coefficient, respectively. Using these relations, we can calculate $M_\mathrm{H\textsc{ii}}$ (in $M_\odot$) from the observed parameters $L(\mathrm{H\alpha})$ (in erg~s$^{-1}$) and $R_S$ (in parsecs) as 

\begin{equation}
M_\mathrm{H\textsc{ii}} = 1.57\times10^{-17}\sqrt{{L(\mathrm{H\alpha})}\times R_S^3}
\end{equation}

We could precisely measure the value of $R_S$ neither for the SSC nor for the ``centre'' because of insufficient spatial resolution, so we estimated it as $FWHM$ of a 2D Gaussian that we used to fit the observed flux distribution of the \Ha broad component for each region. We obtained that $R_s\simeq410$ and 370~pc for the SSC and the ``centre'', respectively. Hence, $M_\mathrm{H\textsc{ii}} \simeq 6.1\times10^6$ and $3.5\times10^6\ M_\odot$ for these regions, and the energy of the outflowing gas should be $E_{kin}=\frac{3}{2}M_\mathrm{H\textsc{ii}}\sigma_\mathrm{gas}^2 = 1.6\times10^{53}$ and $1.7\times10^{53}$~erg.

The mechanical energy produced by SNe could be estimated as in \citet{Tamburro2009}:

\begin{equation}
E_{SNe} = \eta \times SFR \times \epsilon_{SN} \times E_{SN} \times \tau_D,
\end{equation}
where supernovae rate $\eta \simeq 0.01 M_\odot^{-1}$, turbulence decay time $\tau_D \simeq 9.8$~Myr , energy of supernova explosion $E_{SN}\simeq10^{51}$~erg and its fraction converted to the mechanical energy $0<\epsilon<1$. The later value is very uncertain, and here we assume $\epsilon=0.1$ as derived from simulations \citep{Thornton1998}.
From this we obtain $E_{SNe}=4.9\times10^{53}$ and $2.3\times10^{53}$~erg for the SSC and ``centre'' regions respectively that is more then enough to drive the outflow. Note, however, that the age of the young stellar population in the SSC might be younger than the adopted value $\tau_D$ (see Sec.~\ref{sec:sed}), however the available energy will still be  sufficient.
Hence, from this rough estimate, the available mechanical energy only from the SNe feedback is already enough to explain the observed broad underlying components in the SSC and the ``centre'' regions.

The calculation above is reliable assuming that the broad underlying components represent an outflow of the turbulent gas. However, it could be also produced by shock waves resulting from the interaction of stellar winds with surfaces of molecular clouds \citep[e.g.][]{Chu1994, Bresolin2020}. The fact that we do not observe shock excitation there indicated by the BPT diagrams can be explained by a dominant contribution of photoionisation by OB stars. In such a case, the origin of the underlying broad component in the SSC and the ``centre'' still can be easily explained by stellar feedback.

The broad component at the outskirts of \revone{central component} to the east of the SSC has a higher velocity dispersion $\sigma_\mathrm{gas}\sim 70 \kms$, and it is narrower and blue-shifted to the South-East of the SSC. Here its distribution does not show a correlation with any sources of mechanical energy and probably shock waves from from SNe and stellar winds in SSC or from the interaction with the LSB disc penetrating through the lower density ISM could be responsible for these low intensity components.

Hence, from the decomposition of the \Ha line profiles we can draw a conclusion that the velocity dispersion of the ionised gas in the central regions of Ark~18 does not substantially differ from that in the rest of the galaxy, and the broadening of the \Ha line seen in Fig.~\ref{fig:FPI} is caused by a broad underlying component. The latter could be explained by supernovae feedback from the brightest star-forming regions, while additional sources of shock waves may also cause supersonic motions at the outskirts of the \revone{central component}.

\section{\revone{Discussion and Summary}}\label{sec:discuss}

In this paper we analyse the properties of ionised gas, stellar populations and dark matter halo of the low-mass galaxy Ark~18 with an LSB disc that resides in the Eridanus void. We present new observations with the long-slit spectrograph and the scanning Fabry-Perot interferometer at 6-m telescope BTA. These data were accompanied by multi-wavelength archival images and SDSS spectra.
From the analysis presented above we derived the following main features of the Ark~18, based on which we discuss further the possible evolutionary scenarios:

\begin{itemize}
    \item The analysis of internal structure from SDSS images and the kinematics of ionised gas indicate that both the LSB and the central \revone{component} are moderately inclined to each other and can be explained by a pure circular rotation. Feedback-driven non-circular motions are observed as broad underlying components towards the brightest star-forming regions. It is possible that the outer disc is slightly warped. From the rotation curve we \revone{derived the dynamical time (estimated as the period of one revolution at the radius of $\sim40$~arcsec where the curve reaches a plateau)} to be about $300-350$~Myr. 
    \item Almost all emission-line regions are \HII regions ionised by massive stars. The contribution of shocks might be significant only at the outskirt of the \revone{central component} where it contacts the LSB disc. Such behaviour is observed in polar-ring galaxies \citep{Egorov2019}, and probably it might be also expected for smaller relative inclination of the components.
    \item No oxygen abundance gradient is observed in the galaxy. Such a picture is often observed in interacting and merging systems \citep[see, e.g.,][]{Rupke2010,zasov2015,Zasov2019} and points to the efficient ISM mixing.
    \item The \revone{central component} demonstrates the continuous SF history during the large time with a probably constant SFR, and the young population that was formed recently (light-weighted age of the young stars in the ``centre'' is $\sim140$~Myr, while the SSC appears to be much younger). We couldn't obtain reliable estimate of the age of stellar population in LSB disc because of its faintness in UV and IR bands, however the observed SED is consistent with a presence of young stellar population with the age between 60 and 500~Myr.
    \item From the results of {\sc galfit} modelling and spectral and SED fitting we have estimated the ratio of stellar masses of the \revone{central component} and LSB disc to be \revone{at least} $\sim5:1$.
    \item Comparing the obtained values of the stellar mass, oxygen abundance and SFR, we conclude that Ark~18 currently lies on the fundamental mass--metallicity relation (FMR) as constructed by \cite{Curti2020} for the metallicities derived with the $T_e$ method.
    \item Most of current star formation in the galaxy occurs in the two central regions (the SSC and the ``centre'') which contribute about 80 per cent to the total \Ha flux in the galaxy.  
    The relative abundances of N/O, S/O, Ne/O, Ar/O point to their normal chemical evolution, however we observe very uncertain enhancement of these ratios in the ``centre'' region which might be related to a recent accretion of metal poor gas.
    
\item Our estimates of baryonic mass and rotation velocity correspond to non-outlying position of Ark~18 on the baryonic Tully-Fisher relation \citep{tf}. It can indicate that Ark~18 has different formation history in comparison with gas-rich ultra-diffuse galaxies \citep{2019ApJ...883L..33M}, and that it did not require the very inefficient feedback in its formation history. 
    \item Ark~18 is dark matter dominated gas-rich galaxy with dark halo having the radial scale and central density close to those observed in the ``normal'' high surface brightness discy galaxies with the same disc radii.
\end{itemize}

In Ark~18 we observe the central bright circularly rotating \revone{central component} embedded in the extended blue LSB disc with the position and inclination angle differing from that of the central component. The presented comparison of the photometric and kinematic properties of the components suggests the external origin of the LSB structure. Thus, the outer part of Ark~18 could be formed either by cold gas accretion or by a merger. In the case of accretion from a cosmic filament we would expect to see the radial gradient of the metallicity and the low metalicity in the outskirts of Ark~18, since in that case the accreted gas should be metal-poor. We do not observe this phenomenon in Ark~18, instead the oxygen abundance shows flat radial distribution and its value is typical for the Ark~18 luminosity and stellar mass. This suggests that the LSB disc was formed from pre-enriched material at the timescale longer than the dynamical time to allow a good mixing of the ISM to occur.

It is also possible that the gas originates from a gas-rich galaxy that was accreted by the progenitor of Ark~18. Since we do not observe a metallicity gradient, it indicates that either the gas metallicity of the accreted satellite was close to that of the progenitor, or that almost all the gas content of Ark~18 is related to the satellite (it means that the progenitor was a dwarf elliptical galaxy). However, the estimated ages of stellar population from the SED fitting contradict this idea -- the presence of gas is necessary to maintain a constant rate of star formation for long period of time. Also in this case the initial mass of the accreted satellite should be too large because all H~\textsc{i}, the dominant component in Ark~18, must be related to it. 

The scenario in which the satellite mass is several times as low as that of the host is in a better agreement with the presented observational data and is also consistent with our estimate of the ratio of the stellar masses for the central and LSB components. The absence of a significant metallicity gradient suggests that this event occurred earlier than $\sim300$~Myr ago (dynamical timescale for the Ark~18) so the gas had the time to mix. The morphological structure of Ark~18 resembles 
that of some of giant LSBs, with low surface brightness discs outside brighter central region. Thus Ark~18 could be the  lower-mass analogue of gLSBs. The lower mass could be due to more sparse environment of Ark~18, since gLSBs are mostly parts of poor groups and do not belong to voids \citep{Saburova2018}. Mergers are considered as one of the possible scenario for formation of LSB structures in these objects \citep[see, e.g., models proposed in][]{penarrubia2006, zhu2018, Saburova2018}, similar scenario could work also for Ark~18. 

\revone{Ark~18, with its blue LSB features around a redder central part looks quite similar to the nearby galaxy NGC~404. \cite{delRio2004} analyzed H{\sc i} data for NGC~404 and concluded that it is a remnant of a merger with a gas-rich dwarf irregular galaxy $0.5 - 1$~Gyr ago.
We found several more galaxies with the morphology resembling that of Ark~18, including two objects in the same void, UM~40 (with an underlying disc well visible in GALEX data) and PGC3080241.} 


We may also expect an additional recent infall of a small gas cloud of low metallicity to the central part of Ark~18 that diluted the gas and completely flattened the gradient. There are indications that the bright SSC seen near the centre of Ark~18 is a short-lived formation that could be a remnant of such an event. However, a difference between the metallicity of the galaxy and the cloud should be not high since we do not observe any significant effect of the dilution in the relative abundance of different elements. 

Thus, we conclude that Ark~18 could be a result of two encounters. The first one is a dwarf-dwarf merger with an intermediate mass ratio (\revone{at least} 5:1 assuming the \revone{stellar} mass of the LSB disc as the lower limit of the mass \revone{of the secondary component}) that occurred more than 300~Myr ago and led to the formation of the LSB disc. 
However, it is worth mentioning, that our estimate of the initial mass ratio was obtained only for stellar components \revone{and also does not include any constrains on the possible mixing after the merger event}. To derive a more reliable value the spatially resolved \HI data are required.
The second event was a minor merger or an infall of a massive gas cloud and it became responsible for the peculiar inner structure and the ignition of star formation there.

The galaxy is considered isolated at the present epoch, but our results suggest that it had companions in the recent past. Voids may contain groups of galaxies \citep[like NGC~428 group,][]{Egorova2019}, pairs and triplets \citep{J0723,LC1,NVG}, interacting and merging systems \citep{Chengalur17,VGS2013}. In particular, \citet{UGC4277} investigated another void galaxy UGC~4722 that was considered to be one of the most isolated galaxies in the Local Supercluster, though revealing the peculiar morphology. Authors conclude that actually it is a merging system, and the second component, a very gas-rich dwarf, was almost completely destroyed. So the dwarf-dwarf merger scenario for the formation of Ark~18 suggested above does not look improbable despite the low density environment.

\section*{Acknowledgements}

We thank the anonymous referee for constructive comments.
The work is supported by the Russian Science Foundation, grants 19-72-00149 (the analysis of the ionized gas kinematics) and 19-12-00281 (the development of stellar population models and the spectrophotometric fitting technique {\sc NBurst+phot}). The Russian Foundation for Basic Research (project no. 18-32-20120) supported the general analysis of evolutionary scenarios.
OE and KG acknowledge the support from the Foundation of development of theoretical physics and mathematics ``Basis''.
IC's research is supported by the Telescope Data Center at Smithsonian Astrophysical Observatory. 
AS, KG, and IC are also supported by the Interdisciplinary Scientific and Educational School of Moscow University ``Fundamental and Applied Space Research''.
Observations with the 6-m Russian telescope (BTA) were carried out with the financial support of the Ministry of Education and Science of the Russian Federation (agreement no. 14.619.21.0004, project ID RFMEFI61914X0004). The authors thank Dmitry Oparin and Roman Uklein for their assistance in observations with the 6-m telescope, and Simon Pustilnik for useful comments.
This research has made use of the NASA/IPAC Extragalactic Database (NED) which is operated by the Jet Propulsion Laboratory, California Institute of Technology, under the contract with the National
Aeronautics and Space Administration.
The authors acknowledge the spectral and photometric data and the related
information available in the SDSS database used for this study.
Funding for SDSS-III has been provided by the Alfred P. Sloan Foundation,
the Participating Institutions, the National Science Foundation, and the
U.S. Department of Energy Office of Science. The SDSS-III web site is
http://www.sdss3.org/.
Based in part on observations obtained as part of the VISTA Hemisphere Survey, ESO Progam, 179.A-2010 (PI: McMahon)


\section*{DATA AVAILABILITY}
The data underlying this article will be shared on reasonable request to the corresponding author.
The FPI data will be available soon in SIGMA-FPI database (Egorov et al., in preparation).



\bibliographystyle{mnras}
\bibliography{accretion}


\bsp	
\label{lastpage}
\end{document}